\documentclass[journal]{IEEEtran}
\usepackage{amsmath,amsfonts}
\usepackage{algorithmic}
\usepackage{array}
\usepackage{textcomp}
\usepackage{stfloats}
\usepackage{url}
\usepackage{verbatim}
\usepackage{graphicx}
\def\BibTeX{{\rm B\kern-.05em{\sc i\kern-.025em b}\kern-.08em
    T\kern-.1667em\lower.7ex\hbox{E}\kern-.125emX}}
\usepackage{balance}
\usepackage{hyperref}
\usepackage{cite}
\usepackage{tikz}
\usepackage[utf8]{inputenc}
\usepackage{graphicx}
\usepackage{amsmath, amssymb}
\usepackage{enumitem}
\usepackage{siunitx}
\usepackage{pgfplots}
\usepackage{xcolor}
\usepackage{subcaption}
\usepackage{comment}
\usepackage{marvosym}
\usepackage{multirow}
\usepackage{booktabs}
\usepackage[switch]{lineno}

\usepackage[table]{xcolor}

\definecolor{mydarkgreen}{RGB}{0,120,0}
\definecolor{Cyan}{RGB}{0,180,180}
\definecolor{darkCyan}{RGB}{0,90,100}

\makeatletter
\@twocolumntrue
\makeatother

\begin{document}

\title{Advanced Reactance Method - A Novel Approach for Distance Protection in Presence of Grid-Following Inverter-Based-Resources}

\author{Jonathan Loebel,~\IEEEmembership{Graduate Student Member,~IEEE}, Jannik Polster, Johann Jaeger,~\IEEEmembership{Member,~IEEE} \\
	\textcolor{red}{This work has been submitted to the IEEE for possible publication. Copyright may be transferred without notice, after which this version may no longer be accessible.}
	\thanks{J. Loebel and J. Jaeger are with the Institute of Electrical Energy Systems, Friedrich-Alexander-Universität Erlangen-Nürnberg, Erlangen, Germany (e-mail: jonathan.loebel@fau.de)}
	\thanks{J. Polster is with Siemens Energy AG, Erlangen, Germany}
	}

\markboth{ }%
{Loebel \MakeLowercase{\textit{et al.}}: Distance Protection in Presence of Grid-Following Inverter-Based Resources -- Correct Parameterization and Advancement of the Reactance Method}

\maketitle

\begin{abstract}
	Renewable Energy Sources and consequently inverter-based resources (IBRs) are increasing worldwide.
	Grid-following (GFL) IBR control is the state-of-the-art.
	GFL IBRs exhibit different fault behavior compared to classical synchronous generators -- particularly in terms of Fault-Ride-Through (FRT) characteristics, which are critical for grid protection relaying.
	
	Grid protection must work securely, selectively and dependably, regardless of the feeding source.
	As common protection algorithms are designed for fault behavior determined by synchronous generators, their performance is expected to be compromised in the presence of GFL IBRs.
	Distance Protection is a widely used protection algorithm at all voltage levels. 
	Literature shows that the FRT behavior of GFL IBRs impacts the Distance Protection relaying adversely.
	
	This paper proposes the Advanced Reactance Method, a further development of the Classical Reactance Method for Distance Protection in the presence of GFL IBRs.
	The key idea of the proposed method is to make the algorithm independent of the feeding source impedance.
	Therefore, the negative sequence voltage is taken into account along with the negative sequence current.
	This makes the algorithm independent of the FRT behavior of GFL IBRs, even in case of suppression of negative sequence current during unsymmetrical faults.
	Moreover, correct settings for the classical method in the presence of GFL IBRs are presented as a solution for relays without the advanced method.
	
	The novel algorithm is evaluated against the correctly parameterized Classical Reactance Method and the Distance Protection algorithm.
\end{abstract}


	\section{Introduction}
	\label{C:1_Introduction}
	\IEEEPARstart{T}{he} share of renewable energy sources (RES) is increasing rapidly.
	Both, the transmission and the distribution grid are affected.
	Although the majority of installed capacity is in the distribution grid, large projects like High-Voltage Direct Current (HVDC) transmission lines and large battery storages are implemented in the transmission grid \cite{federal_network_agency_eeg_2019,transnetbw_gmbh_ultranet_2023,tennet_tso_gmbh__transnetbw_gmbh_suedlink_2024,tennet_tso_gmbh_suedostlink_2025,NEP2037_2045_2025_2ndDraft}.
	All these technologies are predominantly connected to the grid via inverters.
	Thus, the influence of inverter-based resources (IBRs) on the behavior of the electrical grid increases rapidly.
	The state of the art for modern IBR control is grid-following (GFL).
	These GFL IBRs need an external voltage source for synchronization while the current is set by the control.
	This leads to a completely different behavior, in particular during grid faults, compared to classical synchronous generators\cite{plet_fault_2010}.
	Literature shows that this changed behavior has a negative influence on Distance Protection\cite{hooshyar_distance_2015,brantl_application_2020,davi_case_2023,jia_impact_2019,kleemann_converter_2021,novikov_performance_2019,LoebelDPSP2025}.
	The remote feed problem has proven to be particularly critical, explained in detail by the authors in \cite{loebel_distance_nodate}.
	
	The remote-feed problem also existed in classical, synchronous machine fed, electrical grids \cite{ziegler_digitaler_2008}.
	The problem can be reduced by implementing features such as beveled zone boundaries or load flow compensation.
	Alternatively, an extended Distance Protection algorithm can be used, the so-called Reactance Method.
	This method was developed in several steps from the idea of load flow compensation \cite{takagi_new_1981,takagi_new_1982,takagi_development_1982,eriksson_accurate_1985,dzienis_novel_2014}.
	In \cite{takagi_development_1982} Takagi et al introduce a single sided fault location algorithm based on a comparison of the pre fault condition and faulted network.
	A main idea is introduced: Using the imaginary part of the network equations to eliminate the fault resistance.
	Eriksson et al  introduce a fault current distribution factor in \cite{eriksson_accurate_1985}, that takes sequence networks into account.
	Further they introduce a second important idea: Using representative value of the source impedances of the local and remote end of the line.
	In \cite{dzienis_novel_2014} the approach is developed further, implementing it into classical distance protection algorithms.
	These algorithms were finally implemented in Siemens protection relays \cite{siemens_ag_siprotec_2023}.
	This paper refers to the status of the Reactance Method implemented in these relays as the “Classical Reactance Method”.
	This method was tested in the presence of IBRs in \cite{banaiemoqadam_impact_2023}.
	Several malfunctions of the algorithm are shown, while the IBR control follows classical FRT techniques based on typical grid codes.
	It is found that line-to-line faults (2pF) are significantly more critical than line-to-ground faults (1pF).
	For this reason, this paper refers to 2pF.
	Moreover, in \cite{polster2026faultlocation} the authors of this paper analyze the influence of GFL IBR FRT on the Classical Reactance Method.
	The main findings are that the correct parameterization of the Classical Reactance Method is important.
	Using the correct angle of the internal IBR impedance is much more important than the magnitude.
	Moreover, the GFL IBR must follow FRT with negative sequence currents, as the Classical Reactance Method relies on this and does not work if it is suppressed.
	In general, however, the Classical Reactance Method is capable of outperforming Distance Protection.
	
	The idea is based on a source free negative sequence with a short circuit behind a fixed negative source impedance.
	This assumption is not valid anymore under the influence of GFL IBRs.
	These might suppress negative sequence current or in the case of negative sequence current injection requirements of grid codes, the actual behavior may not be reliable \cite{Nagpal2025}.
	George et al. \cite{George2023} propose an adaptive distance-protection method that estimates the equivalent source impedances at both ends of the line using local measurements.
	However, their method does not explicitly consider the negative-sequence network.
	Paladhi et al. \cite{Paladhi2021,Paladhi2024} address the negative sequence behavior of IBRs.
	In \cite{Paladhi2021}, an algorithm based on local negative-sequence voltage is proposed for cases where local negative-sequence current is unavailable, assuming balanced converter current.
	The method in \cite{Paladhi2024} considers both suppressed and injected negative-sequence current by representing the IBR as an open or closed branch and applying different calculation expressions for each case.
	However, it does not provide a general online procedure for identifying changes in the converter's negative-sequence operating mode.
	This motivates the development of a single-ended distance-protection algorithm based on fault-location principles that accurately represents the negative-sequence network under both suppressed and injected converter-current conditions.
	Ideally, such an algorithm should not require detailed inverter-control information or prior knowledge of the converter's negative-sequence operating mode, nor rely on switching between separately derived equation sets.
	
	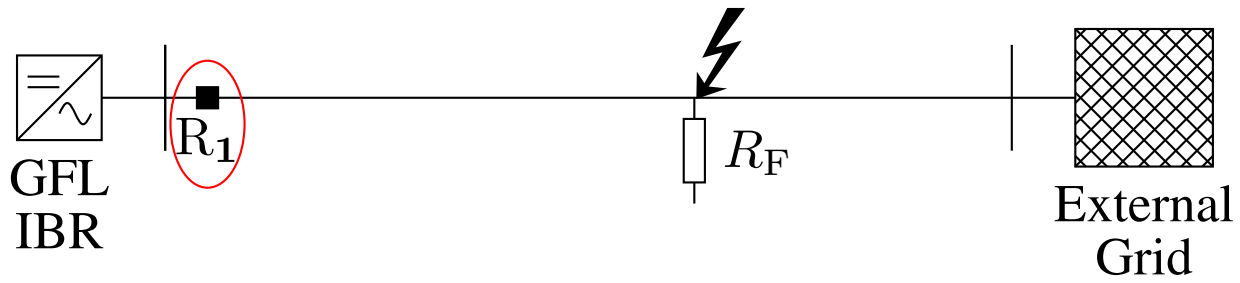
\begin{figure}[!t]
		\centering
		\begin{tikzpicture}[scale=0.7]
			\draw (1.6,-0.4) rectangle (2.4,0.4);
			\draw (1.6,-0.4) -- (2.4,0.4);				
			\draw (2.0,-0.15) sin (2.075,-0.05) cos (2.15,-0.15) sin (2.225,-0.25) cos (2.30,-0.15);
			\draw (1.7, 0.2) -- (2.0, 0.2);
			\draw (1.7, 0.1) -- (2.0, 0.1);
			\node[align=center, inner sep=0pt] at (2,-1) {GFL\\[-2pt]IBR};
			\node[align=center, inner sep=0pt] at (11.6+0.65,-1.25) {External\\[-2pt]Grid};		
			\draw (11.6,-0.65) rectangle (12.9,0.65);		
			\draw (11.6,-0.45) -- (12.7,0.65);
			\draw (11.6,-0.25) -- (12.5,0.65);
			\draw (11.6,-0.05) -- (12.3,0.65);
			\draw (11.6, 0.15) -- (12.1,0.65);
			\draw (11.6, 0.35) -- (11.9,0.65);
			\draw (11.6,-0.65) -- (12.9,0.65);
			\draw (11.8,-0.65) -- (12.9,0.45);
			\draw (12.0,-0.65) -- (12.9,0.25);
			\draw (12.2,-0.65) -- (12.9,0.05);
			\draw (12.4,-0.65) -- (12.9,-0.15);
			\draw (12.6,-0.65) -- (12.9,-0.35);
			\draw (11.6,0.45) -- (12.7,-0.65);
			\draw (11.6,0.25) -- (12.5,-0.65);
			\draw (11.6,0.05) -- (12.3,-0.65);
			\draw (11.6,-0.15) -- (12.1,-0.65);
			\draw (11.6,-0.35) -- (11.9,-0.65);
			\draw (11.6,0.65) -- (12.9,-0.65);
			\draw (11.8,0.65) -- (12.9,-0.45);
			\draw (12.0,0.65) -- (12.9,-0.25);
			\draw (12.2,0.65) -- (12.9,-0.05);
			\draw (12.4,0.65) -- (12.9,0.15);
			\draw (12.6,0.65) -- (12.9,0.35);
			\draw (12.8,-0.65) -- (12.9,-0.55);
			\draw (12.8,0.65) -- (12.9,0.55);
			\draw (11.6,0.55) -- (11.7,0.65);
			\draw (11.6,-0.55) -- (11.7,-0.65);		
			\draw (3, -0.5) -- (3, 0.5);
			\draw (3, -0.5) -- (3, 0.5);
			\draw (11, -0.5) -- (11, 0.5);
			\draw (2.4, 0) -- (11.6, 0);
			\node at (8.25,0.41) {\Huge\Lightning};
			\draw (8, 0) -- (8, -1);
			\draw[fill=white, draw=black] (7.9,-0.2) rectangle (8.1,-0.8);
			\node at (8.6, -0.52) {$R_{\mathrm{F}}$};
			\draw[fill=black, draw=black] (3.3,-0.1) rectangle (3.5,0.1);
			\node at (3.4, -0.4) {$\mathrm{R}_{\mathbf{1}}$};
			\draw[red] (3.4,-0.25) ellipse (0.35 and 0.6);
		\end{tikzpicture}
		\caption{The remote feed problem: A resistive grid fault fed from both sides. The distance protection relay on the GFL IBR side shows high measurement deviations.}
		\label{fig:1_Remote Feed Problem}
	\end{figure}
	
	This paper presents the so called "Advanced Reactance Method", a further developed protection algorithm, that is based on the Classical Reactance Method.
	A new substitution of the fault current is introduced, using not only the negative sequence current but also voltage.
	The new algorithm is independent of the negative sequence internal source and can therefore also be used if the GFL IBR FRT does not include negative sequence current injection.
	
	The paper is structured as follows.
	In \ref{C:2_Problem Statement} the remote feed problem is briefly summarized.
	In addition, the Classical Reactance Method is explained, as well as the influence of GFL IBR on it.
	For the absence of negative sequence current during FRT the Classical Reactance Method no longer works.
	Therefore, in \ref{C:3_New Fault Current Substitution} the Advanced Reactance Method is presented.
	As the novel method is not implemented in all protection relays, the correct parameterization of the Classical Reactance Method in the presence of GFL IBRs is discussed in \ref{C:31_Parameterization of the Classical Reactance Method in the Presence of GFL IBR}.
	Subsequently, the Distance Protection, Classical Reactance Method and Advanced Reactance Method are compared in a test grid in \ref{C:4_Application and Parameterization of the Reactance Method for GFL IBRs}.
	Different FRT techniques of GFL IBRs are considered, as well as a changing internal source.
	Moreover, the proposed parameterization of the Classical Reactance Method is evaluated.
	A summary and outlook are given in \ref{C:5_Conclusion} and \ref{C:6_Outlook}.
	
	\section{Problem Statement}
	\label{C:2_Problem Statement}
	
	\subsection{Remote Feed Problem in Distance Protection}
	\label{C:21_Remote Feed Problem}
	
	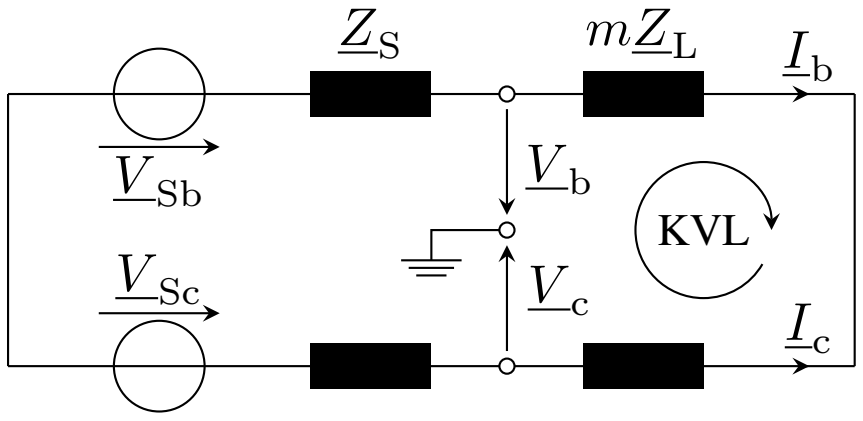
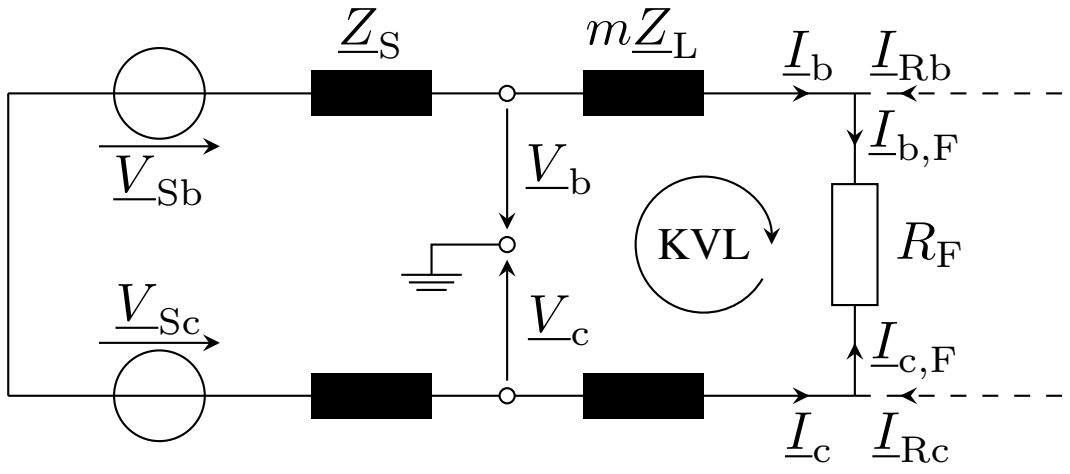
\begin{figure}[t!]
		\centering
		\captionsetup[subfigure]{skip=0pt}
		\setlength{\abovecaptionskip}{2pt}
		\setlength{\belowcaptionskip}{0pt}
		\begin{subfigure}{1.0\columnwidth}
			\centering
			\begin{tikzpicture}
				\draw (-4,0) -- (1.6,0);
				\draw (-4,-1.8) -- (1.6,-1.8);
				\draw (-4,0) -- (-4,-1.8);
				\draw (1.6,0) -- (1.6, -1.8);
				\draw[->, stealth-] (1.3,0) -- ++(-0.01, 0);
				\draw[->, stealth-] (1.3,-1.8) -- ++(-0.01, 0);
				\draw(-3,0) circle [radius=0.3cm];
				\draw(-3,-1.8) circle [radius=0.3cm];
				\draw[<-, -stealth] (-3.4,-0.35) -- (-2.6, -0.35);
				\node at (-3, -0.575) {$\underline{V}_{\mathrm{Sb}}$};
				\draw[<-, -stealth] (-3.4,-1.45) -- (-2.6, -1.45);
				\node at (-3, -1.225) {$\underline{V}_{\mathrm{Sc}}$};
				\node at (-1.6, 0.4) {$\underline{Z}_{\mathrm{S}}$};
				\draw[fill=white, draw=black] (-0.7,0) circle [radius=0.05cm];
				\draw[fill=white, draw=black] (-0.7,-1.8) circle [radius=0.05cm];
				\draw[fill=white, draw=black] (-0.7,-0.9) circle [radius=0.05cm];
				\draw (-0.75,-0.9) -- (-1.2,-0.9) -- (-1.2,-1.1);
				\draw (-1.4,-1.1) -- (-1.0,-1.1);
				\draw (-1.35,-1.15) -- (-1.05,-1.15);
				\draw (-1.3,-1.2) -- (-1.1,-1.2);
				\draw[->, stealth-] (-0.7,-0.8) -- (-0.7, -0.1);
				\draw[<-, -stealth] (-0.7,-1.7) -- (-0.7, -1.0);
				\node at (-0.35, -0.5) {$\underline{V}_{\mathrm{b}}$};
				\node at (-0.35, -1.3) {$\underline{V}_{\mathrm{c}}$};
				\fill[black] (-2,0.15) rectangle (-1.2,-0.15);
				\fill[black] (-0.2,0.15) rectangle (0.6,-0.15);
				\fill[black] (-2,-1.65) rectangle (-1.2,-1.95);
				\fill[black] (-0.2,-1.65) rectangle (0.6,-1.95);
				\node at (0.2, -2.25) {\textcolor{white}{$\underline{Z}_{\mathrm{g}}$}};
				\node at (0.2, 0.4) {$m\underline{Z}_{\mathrm{L}}$};
				\node at (1.3, 0.25) {$\underline{I}_{\mathrm{b}}$};
				\node at (1.3, -1.55) {$\underline{I}_{\mathrm{c}}$};
				\draw[->, stealth-] (0.05,-0.9) ++(0:1cm) arc[start angle=0, end angle=330, radius=0.45cm];
				\node at (0.6,-0.9) {\scalebox{0.85}{KVL}};
			\end{tikzpicture}
			\caption{}
			\label{fig:2.11:Assumption for 2pF algorithm in distance protection}
		\end{subfigure}
		\hfill
		\begin{subfigure}{1.0\columnwidth}
			\centering
			\begin{tikzpicture}
				\draw (-4,0) -- (1.6,0);
				\draw (-4,-2.0) -- (1.6,-2.0);
				\draw (-4,0) -- (-4,-2.0);
				\draw (1.6,0) -- (1.6, -2.0);
				\draw[->, stealth-] (1.3,0) -- ++(-0.01, 0);
				\draw[->, stealth-, black] (1.9,0) -- ++(0.01, 0);
				\draw[->, stealth-] (1.3,-2.0) -- ++(-0.01, 0);
				\draw[->, stealth-, black] (1.9,-2.0) -- ++(0.01, 0);
				\draw[->, stealth-] (1.6,-0.35) -- ++(0, 0.01);
				\draw[->, stealth-] (1.6,-1.65) -- ++(0, -0.01);
				\draw(-3,0) circle [radius=0.3cm];
				\draw(-3,-2.0) circle [radius=0.3cm];
				\draw[fill=white, draw=black] (-0.7,0) circle [radius=0.05cm];
				\draw[fill=white, draw=black] (-0.7,-2.0) circle [radius=0.05cm];
				\draw[<-, -stealth] (-3.4,-0.35) -- (-2.6, -0.35);
				\node at (-3, -0.575) {$\underline{V}_{\mathrm{Sb}}$};
				\draw[<-, -stealth] (-3.4,-1.65) -- (-2.6, -1.65);
				\node at (-3, -1.425) {$\underline{V}_{\mathrm{Sc}}$};
				\node at (-1.6, 0.4) {$\underline{Z}_{\mathrm{S}}$};
				\fill[black] (-2,0.15) rectangle (-1.2,-0.15);
				\fill[black] (-0.2,0.15) rectangle (0.6,-0.15);
				\fill[black] (-2,-1.85) rectangle (-1.2,-2.15);
				\fill[black] (-0.2,-1.85) rectangle (0.6,-2.15);
				\draw[fill=white, draw=black] (1.45,-1.4) rectangle (1.75,-0.6);
				\draw[black, dashed] (1.6,0) -- (3,0);
				\draw[black, dashed] (1.6,-2.0) -- (3,-2.0);
				\draw[fill=white, draw=black] (-0.7,-1.0) circle [radius=0.05cm];
				\draw (-0.75,-1.0) -- (-1.2,-1.0) -- (-1.2,-1.2);
				\draw (-1.4,-1.2) -- (-1.0,-1.2);
				\draw (-1.35,-1.25) -- (-1.05,-1.25);
				\draw (-1.3,-1.3) -- (-1.1,-1.3);
				\node at (0.2, -2.45) {\textcolor{white}{$\underline{Z}_{\mathrm{g}}$}};
				\node at (0.2, 0.4) {$m\underline{Z}_{\mathrm{L}}$};
				\node at (1.3, 0.25) {$\underline{I}_{\mathrm{b}}$};
				\node at (1.3, -2.275) {$\underline{I}_{\mathrm{c}}$};
				\node[black] at (1.98, 0.25) {$\underline{I}_{\mathrm{Rb}}$};
				\node[black] at (1.98, -2.275) {$\underline{I}_{\mathrm{Rc}}$};
				\draw[->, stealth-] (-0.7,-0.9) -- (-0.7, -0.1);
				\draw[<-, -stealth] (-0.7,-1.9) -- (-0.7, -1.1);
				\node at (-0.35, -0.5) {$\underline{V}_{\mathrm{b}}$};
				\node at (-0.35, -1.5) {$\underline{V}_{\mathrm{c}}$};
				\node at (2.1, -1.0) {$R_{\mathrm{F}}$};
				\node at (2.0, -0.3) {$\underline{I}_{\mathrm{b,F}}$};
				\node at (2.0, -1.7) {$\underline{I}_{\mathrm{c,F}}$};
				\draw[->, stealth-] (0.05,-1.0) ++(0:1cm) arc[start angle=0, end angle=330, radius=0.45cm];
				\node at (0.6,-1.0) {\scalebox{0.85}{KVL}};
			\end{tikzpicture}
			\caption{}
			\label{fig:2.12:2pF with fault resistance and remote feed}
		\end{subfigure}
		\caption{Equivalent circuit for 2pF (a) without fault resistance and remote feed as initial assumption for Distance Protection, (b) with fault resistance and remote feed}
		\label{fig:2.1:2pF in Distance Protection}
	\end{figure}
	The Distance Protection algorithm for 2pF between phase b and c is based on the equivalent circuit shown in Fig.~\ref{fig:2.11:Assumption for 2pF algorithm in distance protection}.
	Based on the Kirchhoff's Voltage Law (KVL), the line impedance to the fault location $m\underline{Z}_\mathrm{L}$ can be calculated from the measured phase-to-ground voltages and phase currents, shown in \eqref{eq:2.1:LL Fault}.
	\begin{equation}
		\underline{Z}_{\mathrm{meas,bc}} = \frac{\underline{V}_{\mathrm{b}}-{\underline{V}_{\mathrm{c}}}}{\underline{I}_{\mathrm{b}}-{\underline{I}_{\mathrm{c}}}} = m\underline{Z}_{\mathrm{L}}
		\label{eq:2.1:LL Fault}
	\end{equation}
	The measured impedance is compared to the known specific impedance values of the transmission line.
	The fault location $m$ can thus be calculated and a selective trip decision is made.
	In this initial situation, it is assumed that the fault has no fault resistance $R_\mathrm{F}$.
	If this assumption is incorrect, a measurement error occurs.
	This is in purely resistive direction, whereby the reactance value is taken into account for distance calculation, as this value is not affected.
	
	If the resistive fault is additionally fed from the remote end (R), the equivalent circuit changes further to Fig.~\ref{fig:2.12:2pF with fault resistance and remote feed}.
	As the basic measurement algorithm in \eqref{eq:2.1:LL Fault} stays the same, the unknown remote currents may lead to a measurement deviation of both the resistance and reactance values.
	The complex measurement error is defined by the local and remote currents, shown in \eqref{eq:2.2:LL Fault Remote feed}.
	\begin{equation}
		\begin{split}
			&\underline{Z}_{\mathrm{meas,bc}} = \frac{\underline{V}_{\mathrm{b}}-{\underline{V}_{\mathrm{c}}}}{\underline{I}_{\mathrm{b}}-{\underline{I}_{\mathrm{c}}}} = m\underline{Z}_{\mathrm{L}} + \frac{R_\mathrm{F}}{2} \frac{\underline{I}_\mathrm{b}+\underline{I}_\mathrm{Rb}-\underline{I}_\mathrm{c}-\underline{I}_\mathrm{Rc}}{\underline{I}_{\mathrm{b}}-{\underline{I}_{\mathrm{c}}}}
			\\
			&= m\underline{Z}_{\mathrm{L}} + \frac{R_\mathrm{F}}{2} \frac{\underline{I}_\mathrm{b,F}-\underline{I}_\mathrm{c,F}}{\underline{I}_{\mathrm{b}}-{\underline{I}_{\mathrm{c}}}}
			= m\underline{Z}_{\mathrm{L}} + R_\mathrm{F} \frac{\underline{I}_\mathrm{F}}{\underline{I}_{\mathrm{b}}-{\underline{I}_{\mathrm{c}}}}
		\end{split}
		\label{eq:2.2:LL Fault Remote feed}
	\end{equation}
	
	As a result, also the reactance is no longer a reliable indicator for measuring the distance to the fault.
	In classical grids, this error in the reactance value was a dynamic problem only in specific grid configurations.
	The measurement error was mostly still in resistance direction.
	Previous research shows that with GFL IBRs it is a steady state problem, due to the current source behavior.
	In \cite{loebel_distance_nodate} the authors discuss this effect in great detail.
	
	\subsection{Classical Reactance Method}
	\label{C:22_Reactance Method}
\begin{figure}[!t]
		\centering
		\begin{tikzpicture}[scale=0.57]
			\draw (0,0) -- (12,0);
			\draw (0,-3) -- (12,-3);
			\draw (0,0) -- (0,-3);
			\draw (12,0) -- (12, -3);
			\draw(0,-1.5) circle [radius=0.4];
			\draw(12,-1.5) circle [radius=0.4];
			\fill[black] (0.5,0.175) rectangle (1.5,-0.175);
			\fill[black] (10.5,0.175) rectangle (11.5,-0.175);
			\node[scale=0.9, transform shape] at (3, 0.25) {$\mathbf{B_1^+}$};
			\draw[fill=white, draw=black] (3,0) circle [radius=0.05];
			\draw[fill=white, draw=black] (3,-3) circle [radius=0.05];
			\node[scale=0.9, transform shape] at (9, 0.25) {$\mathbf{B_2^+}$};
			\draw[fill=white, draw=black] (9,0) circle [radius=0.05];
			\draw[fill=white, draw=black] (9,-3) circle [radius=0.05];
			\fill[black] (4,0.175) rectangle (5,-0.175);
			\fill[black] (7,0.175) rectangle (8,-0.175);
			\draw (6,0) -- (6,1) -- (14,1) -- (14,-4) -- (6,-4) --(6, -5);
			\draw (6,-8) -- (6,-9) -- (-1,-9) -- (-1,-3.5) -- (6,-3.5) --(6, -3);
			\draw[fill=white, draw=black] (13.825,-0.25) rectangle (14.175,-1.25);
			\draw[fill=white, draw=black] (13.825,-1.75) rectangle (14.175,-2.75);
			\node at (1, 0.45) {\scriptsize$\underline{Z}_{\mathrm{B1}}$};
			\node at (11, 0.45) {\scriptsize$\underline{Z}_{\mathrm{B2}}$};
			\node at (4.5, 0.45) {\scriptsize$m\underline{Z}_{\mathrm{L}}$};
			\node at (7.5, 0.45) {\scriptsize$(1-m)\underline{Z}_{\mathrm{L}}$};
			\node at (14.55, -0.75) {\scriptsize$\frac{R_{\mathrm{F}}}{2}$};
			\node at (14.55, -2.25) {\scriptsize$\frac{R_{\mathrm{F}}}{2}$};
			\draw[->, stealth-] (2.25,0) -- ++(-0.01, 0);
			\node at (2.25, 0.4) {\scriptsize$\underline{I}^+_{\mathrm{B1}}$};
			\draw[->, stealth-] (9.75,0) -- ++(0.01, 0);
			\node at (9.75, 0.4) {\scriptsize$\underline{I}^+_{\mathrm{B2}}$};
			\draw[->, stealth-] (14,0.4) -- ++(0, 0.01);
			\node at (13.65, 0.5) {\scriptsize$\underline{I}^+_{\mathrm{F}}$};
			\draw[->, stealth-] (14,-3.4) -- ++(0, -0.01);
			\node at (13.65, -3.4) {\scriptsize$\underline{I}^-_{\mathrm{F}}$};
			\draw[->, stealth-] (3,-2.8) -- ++(0, 2.6);
			\node at (3.6, -1.5) {\scriptsize$\underline{V}^+_{\mathrm{B1}}$};
			\draw[->, stealth-] (4.9,-0.3) -- ++(-0.8, 0);
			\node at (4.5, -0.65) {\scriptsize$\underline{V}^+_{\mathrm{mZL}}$};
			\draw[->, stealth-] (7.1,-0.3) -- ++(0.8, 0);
			\node at (7.5, -0.65) {\scriptsize$\underline{V}^+_{\mathrm{(1-m)ZL}}$};
			\draw[->, stealth-] (1.4,-0.3) -- ++(-0.8, 0);
			\node at (1, -0.65) {\scriptsize$\underline{V}^+_{\mathrm{Z,B1}}$};
			\draw[->, stealth-] (10.6,-0.3) -- ++(0.8, 0);
			\node at (11, -0.65) {\scriptsize$\underline{V}^+_{\mathrm{Z,B2}}$};
			\draw[->, stealth-] (9,-2.8) -- ++(0, 2.6);
			\node at (9.6, -1.5) {\scriptsize$\underline{V}^+_{\mathrm{B2}}$};
			\draw[->, stealth-] (0.5,-1.9) -- ++(0, 0.8);
			\node at (1.1, -1.5) {\scriptsize$\underline{V}^{*+}_{\mathrm{B1}}$};
			\draw[->, stealth-] (12.5,-1.9) -- ++(0, 0.8);
			\node at (13.1, -1.5) {\scriptsize$\underline{V}^{*+}_{\mathrm{B2}}$};
			\draw (0,-5) -- (12,-5);
			\draw (0,-8) -- (12,-8);
			\draw (0,-5) -- (0,-8);
			\draw (12,-5) -- (12, -8);
			\fill[black] (0.5,-4.825) rectangle (1.5,-5.175);
			\fill[black] (10.5,-4.825) rectangle (11.5,-5.175);
			\node[scale=0.9, transform shape] at (3, -4.75) {$\mathbf{B_1^-}$};
			\draw[fill=white, draw=black] (3,-5) circle [radius=0.05];
			\draw[fill=white, draw=black] (3,-8) circle [radius=0.05];
			\node[scale=0.9, transform shape] at (9, -4.75) {$\mathbf{B_2^-}$};
			\draw[fill=white, draw=black] (9,-5) circle [radius=0.05];
			\draw[fill=white, draw=black] (9,-8) circle [radius=0.05];
			\fill[black] (4,-4.825) rectangle (5,-5.175);
			\fill[black] (7,-4.825) rectangle (8,-5.175);
			\node at (1, -4.55) {\scriptsize$\underline{Z}_{\mathrm{B1}}$};
			\node at (11, -4.55) {\scriptsize$\underline{Z}_{\mathrm{B2}}$};
			\node at (4.5, -4.55) {\scriptsize$m\underline{Z}_{\mathrm{L}}$};
			\node at (7.5, -4.55) {\scriptsize$(1-m)\underline{Z}_{\mathrm{L}}$};
			\draw[->, stealth-] (2.25,-5) -- ++(-0.01, 0);
			\node at (2.25, -4.6) {\scriptsize$\underline{I}^-_{\mathrm{B1}}$};
			\draw[->, stealth-] (9.75,-5) -- ++(0.01, 0);
			\node at (9.75, -4.6) {\scriptsize$\underline{I}^-_{\mathrm{B2}}$};
			\draw[->, stealth-] (3,-7.8) -- ++(0, 2.6);
			\node at (3.6, -6.5) {\scriptsize$\underline{V}^-_{\mathrm{B1}}$};
			\draw[->, stealth-] (4.9,-5.3) -- ++(-0.8, 0);
			\node at (4.5, -5.65) {\scriptsize$\underline{V}^-_{\mathrm{mZL}}$};
			\draw[->, stealth-] (1.4,-5.3) -- ++(-0.8, 0);
			\node at (1, -5.65) {\scriptsize$\underline{V}^-_{\mathrm{Z,B1}}$};
			\draw[->, stealth-] (7.1,-5.3) -- ++(0.8, 0);
			\node at (7.5, -5.65) {\scriptsize$\underline{V}^-_{\mathrm{(1-m)ZL}}$};
			\draw[->, stealth-] (10.6,-5.3) -- ++(0.8, 0);
			\node at (11, -5.65) {\scriptsize$\underline{V}^-_{\mathrm{Z,B2}}$};
			\draw[->, stealth-] (9,-7.8) -- ++(0, 2.6);
			\node at (9.6, -6.5) {\scriptsize$\underline{V}^-_{\mathrm{B2}}$};
		\end{tikzpicture}
		\caption{Equivalent circuit of 2pF with fault resistance and remote feed in symmetrical components}
		\label{fig:2.2:2pF in SymC}
	\end{figure}
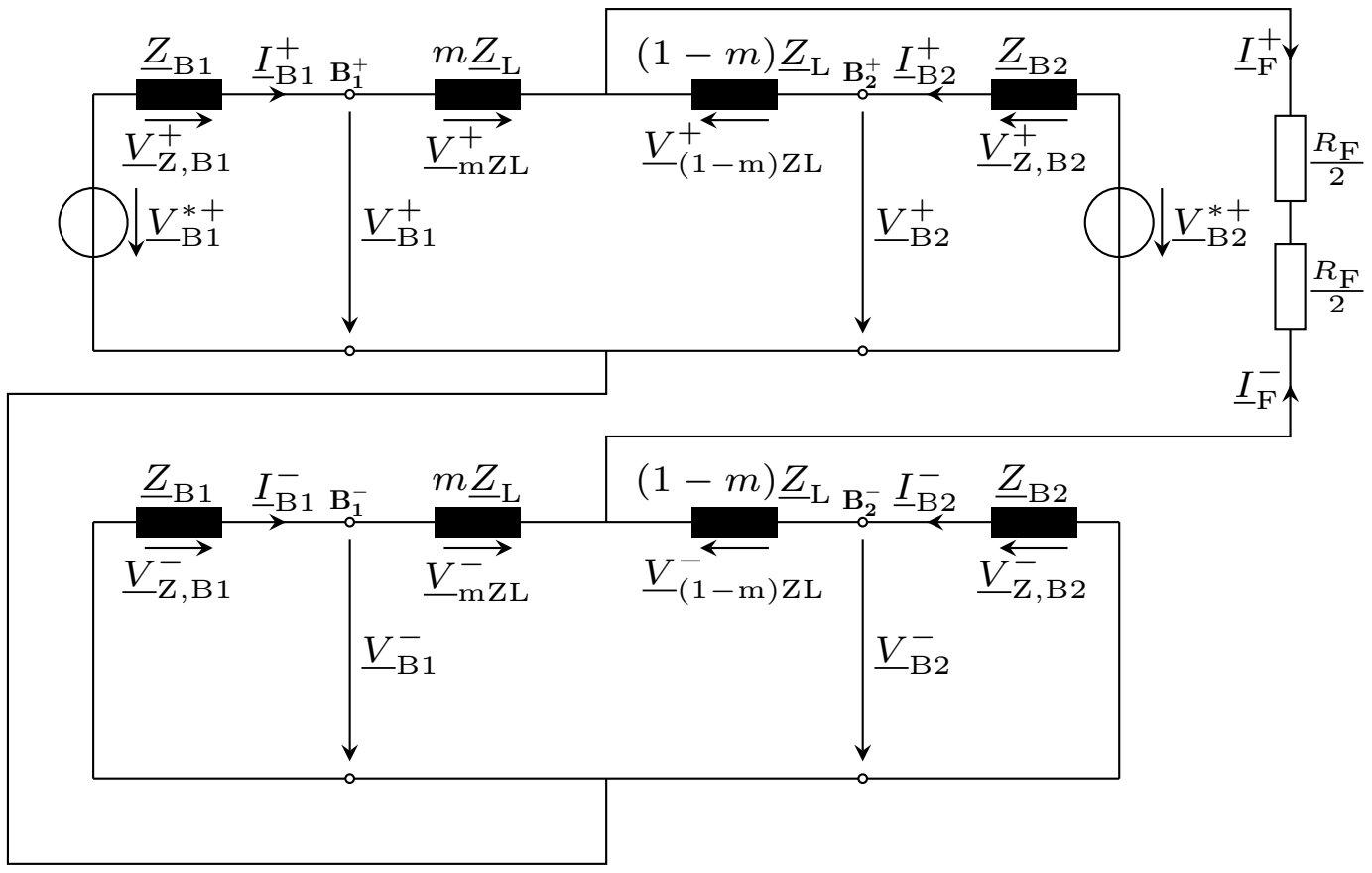

	The Classical Reactance Method offers a solution for the remote-feed problem in classical, synchronous machine fed grids.
	It is based on the equivalent circuit of the 2pF with remote feed in symmetrical components, shown in Fig. \ref{fig:2.2:2pF in SymC}.
	The positive sequence is denoted as \mbox{'+'}, the negative sequence \mbox{'-'}, respectively.
	A transmission line with the impedance $\underline{Z}_\mathrm{L}$ is connected to two buses, B\textsubscript{1} and B\textsubscript{2}.
	On each side, a grid is represented by a voltage source in the positive sequence and a short-circuit in the negative sequence with a corresponding source impedance $\underline{Z}_\mathrm{B1}$ and $\underline{Z}_\mathrm{B2}$.
	The fault location $m$ defines where the positive sequence is coupled with the negative sequence via the fault resistance $R_\mathrm{F}$.
	
	The concept of the Classical Reactance Method is to estimate the fault current and thereby eliminate the measurement error.
	The absence of sources in the negative sequence is used.
	Moreover, the unknown fault resistance is removed by calculating the imaginary part.
	
	The algorithm is derived in the following for a relay at B\textsubscript{1}.
	This is then used as the basis for defining the problem in the presence of GFL IBRs and adapting the algorithm in the next chapter.	
	The fault current between phase b and c is substituted, assuming that the positive sequence fault current corresponds to the opposite negative sequence fault current, shown in \eqref{eq:2_Fault current in abc}.
	\begin{equation}
		\begin{split}
			\underline{I}_\mathrm{F} &= \frac{1}{2}\left( \underline{I}_\mathrm{b,F} - \underline{I}_\mathrm{c,F}\right) 
			\\
			&=\frac{1}{2}\left[  \left( \underline{I}^0_\mathrm{F} +  \underline{a}^2 \underline{I}^+_\mathrm{F} + \underline{a}\underline{I}^-_\mathrm{F}\right)  - \left(  \underline{I}^0_\mathrm{F} +  \underline{a} \underline{I}^+_\mathrm{F} + \underline{a}^2 \underline{I}^-_\mathrm{F} \right) \right] 
			\\
			&=\frac{1}{2}\left( -\underline{a}^2 \underline{I}^-_\mathrm{F} + \underline{a} \underline{I}^-_\mathrm{F} + \underline{a}\underline{I}^-_\mathrm{F} - \underline{a}^2 \underline{I}^-_\mathrm{F}\right)
			\\
			&=\left(\underline{a}-\underline{a}^2\right) \underline{I}^-_\mathrm{F}
		\end{split}
		\label{eq:2_Fault current in abc}
	\end{equation}
	The current distribution factor $\underline{d}^-$ is introduced to represent the ratio of the local negative sequence current to the negative sequence fault current, shown in \eqref{eq:2_I_B1 to I_F}.
	\begin{equation}
		\frac{\underline{I}^-_{\mathrm{B1}}}{\underline{I}^-_\mathrm{F}} = \frac{\underline{Z}_\mathrm{B2}+(1-m)\underline{Z}_\mathrm{L}}{\underline{Z}_\mathrm{B1} + \underline{Z}_\mathrm{B2}+\underline{Z}_\mathrm{L}} = \underline{d}^- = d^- \cdot e^{j\varphi^-_d}
		\label{eq:2_I_B1 to I_F}
	\end{equation}
	For known impedances, the fault current can be calculated from the locally measured negative-sequence current using the factor.
	The grid and line impedances are assumed to be known.
	Thus, only the fault location $m$ is unknown.
	It is set to the zone boundary of the Distance Protection.
	Consequently, accuracy is highest at this critical point.
	The estimation of the fault current based on the locally measured current is now integrated into the distance measurement algorithm.
	First, substituting \eqref{eq:2_I_B1 to I_F} into \eqref{eq:2_Fault current in abc} results in \eqref{eq:2_Fault current I_subst}.
	The negative sequence substituted current $\underline{I}^-_\mathrm{subst}$ is introduced.
	This current is used to estimate the fault current.
	\begin{equation}
		\begin{split}
			\underline{I}_\mathrm{F} &= \frac{1}{\underline{d}^-} \left(\underline{a}-\underline{a}^2\right)\cdot\underline{I}^-_\mathrm{B1}
			\\
			&= \frac{1}{\underline{d}^-} \cdot \underline{I}^-_\mathrm{subst}
		\end{split}
		\label{eq:2_Fault current I_subst}
	\end{equation}
	The fault current in \eqref{eq:2_Fault current I_subst} is now substituted into \eqref{eq:2.2:LL Fault Remote feed} resulting in \eqref{eq:2_Loop_Equation}.
	\begin{equation}
		\begin{split}
			\underline{V}_\mathrm{b} - \underline{V}_\mathrm{c} 
			&= m\underline{Z}_\mathrm{L} \left(\underline{I}_\mathrm{b} -\underline{I}_\mathrm{c} \right) + R_\mathrm{F} \cdot \frac{1}{\underline{d}^-} \underline{I}^-_\mathrm{subst}
			\\ &= m\underline{Z}_\mathrm{L} \left(\underline{I}_\mathrm{b} -\underline{I}_\mathrm{c} \right) + R_\mathrm{F} \cdot \frac{1}{d^-} e^{-j\varphi^-_d} \underline{I}^-_\mathrm{subst}
		\end{split}
		\label{eq:2_Loop_Equation}
	\end{equation}
	The fault resistance is eliminated by converting the right-hand side of the equation purely real.
	This is done by multiplying \eqref{eq:2_Loop_Equation} by the negative angle of the current distribution factor and the complex conjugate of the substituted current in \eqref{eq:2_Loop_Equation_2}.
	\begin{equation}
		\begin{split}
			&\left( \underline{V}_\mathrm{b} - \underline{V}_\mathrm{c} \right) e^{j\varphi^-_d} \overline{\underline{I}}^-_\mathrm{subst}
			\\&= m\underline{Z}_\mathrm{L} \left(\underline{I}_\mathrm{b} -\underline{I}_\mathrm{c} \right) e^{j\varphi^-_d} \overline{\underline{I}}^-_\mathrm{subst}
			+ R_\mathrm{F} \cdot \frac{1}{d^-} \left( I^-_\mathrm{subst}\right) ^2
		\end{split}
		\label{eq:2_Loop_Equation_2}
	\end{equation}
	Calculating the imaginary part of both sides results in \eqref{eq:2_Loop_Equation_Imag_3}.
	\begin{equation}
		\begin{split}
			&\Im \left\{ \left( \underline{V}_\mathrm{b} - \underline{V}_\mathrm{c}\right) e^{j\varphi^-_d} \overline{\underline{I}}^{-}_\mathrm{subst} \right\} =
			\\&\Im \left\{ m\underline{Z}_\mathrm{L} \left( \underline{I}_\mathrm{b} - \underline{I}_\mathrm{c}\right) e^{j\varphi^-_d}\overline{\underline{I}}^{-}_\mathrm{subst} \right\}
			\\
			&= mZ_\mathrm{L} \cdot \Im \left\{ e^{j\varphi_{\mathrm{\underline{Z}L}}} \left( \underline{I}_\mathrm{b} - \underline{I}_\mathrm{c}\right) e^{j\varphi^-_d} \overline{\underline{I}}^{-}_\mathrm{subst} \right\}
		\end{split}
		\label{eq:2_Loop_Equation_Imag_3}
	\end{equation}
	Calculating the reactance by the known line angle results in the Classical Reactance Method algorithm in \eqref{eq:2_Loop_Equation_Imag_4}.
	\begin{equation}
		X_\mathrm{meas} = \frac{\Im \left\{ \left( \underline{V}_\mathrm{b} - \underline{V}_\mathrm{c}\right) e^{j\varphi^-_d} \overline{\underline{I}}^{-}_\mathrm{subst} \right\} }{
			\Im \left\{ e^{j\varphi_{\mathrm{\underline{Z}L}}} \left( \underline{I}_\mathrm{b} - \underline{I}_\mathrm{c}\right) e^{j\varphi^-_d} \overline{\underline{I}}^{-}_\mathrm{subst} \right\} } \sin\left(\varphi_\mathrm{L} \right) 
		\label{eq:2_Loop_Equation_Imag_4}
	\end{equation}
	If the assumed impedances and fault location correspond to reality, the measured reactance is equal to the line reactance up to the fault location.
	
	\subsection{Influence of Grid-Following Inverter Based Resource Feed}
	\label{C:23_Influence of Grid-Following Inverter Based Resource Feed}
	\begin{figure}[!t]
		\centering
		\begin{subfigure}{0.15\textwidth}
			\centering
			\begin{tikzpicture}
				\draw (1,0) -- (2.5,0);
				\draw (1,-2) -- (2.5,-2);
				\draw (1,0) -- (1,-0.6);
				\draw (1,-1.4) -- (1,-2);
				\draw[fill=white, draw=black] (2.5,0) circle [radius=0.05];
				\draw[fill=white, draw=black] (2.5,-2) circle [radius=0.05];
				\draw[fill=white, draw=black] (1,-0.6) circle [radius=0.05];
				\draw[fill=white, draw=black] (1,-1.4) circle [radius=0.05];
				\draw[color=white] (1,-1) circle [radius=0.4];
			\end{tikzpicture}
			\caption{}
			\label{fig:2_GFL IBR NS_a}
		\end{subfigure}
		\begin{subfigure}{0.15\textwidth}
			\centering
			\begin{tikzpicture}
				\draw (1,0) -- (2.5,0);
				\draw (1,-2) -- (2.5,-2);
				\draw (1,0) -- (1,-0.6);
				\draw (1,-0.6) -- (1,-1.4);
				\draw (1,-1.4) -- (1,-2);
				\draw[fill=white, draw=black] (2.5,0) circle [radius=0.05];
				\draw[fill=white, draw=black] (2.5,-2) circle [radius=0.05];
				\fill[black] (1.3,-0.15) rectangle (2.2,0.15);
				\draw[color=white] (1,-1) circle [radius=0.4];
			\end{tikzpicture}
			\caption{}
			\label{fig:2_GFL IBR NS_b}
		\end{subfigure}
		\begin{subfigure}{0.15\textwidth}
			\centering
			\begin{tikzpicture}
				\draw (1,0) -- (2.5,0);
				\draw (1,-2) -- (2.5,-2);
				\draw (1,0) -- (1,-0.6);
				\draw (1,-1.4) -- (1,-2);
				\draw[fill=white, draw=black] (2.5,0) circle [radius=0.05];
				\draw[fill=white, draw=black] (2.5,-2) circle [radius=0.05];
				\fill[black] (1.3,-0.15) rectangle (2.2,0.15);
				\draw (1,-1) circle [radius=0.4];
				\draw (0.6,-1) -- (1.4, -1);
			\end{tikzpicture}
			\caption{}
			\label{fig:2_GFL IBR NS_c}
		\end{subfigure}
		\caption{Typical negative sequence behavior by GFL IBR during FRT}
		\label{fig:2_GFL IBR NS}
	\end{figure}
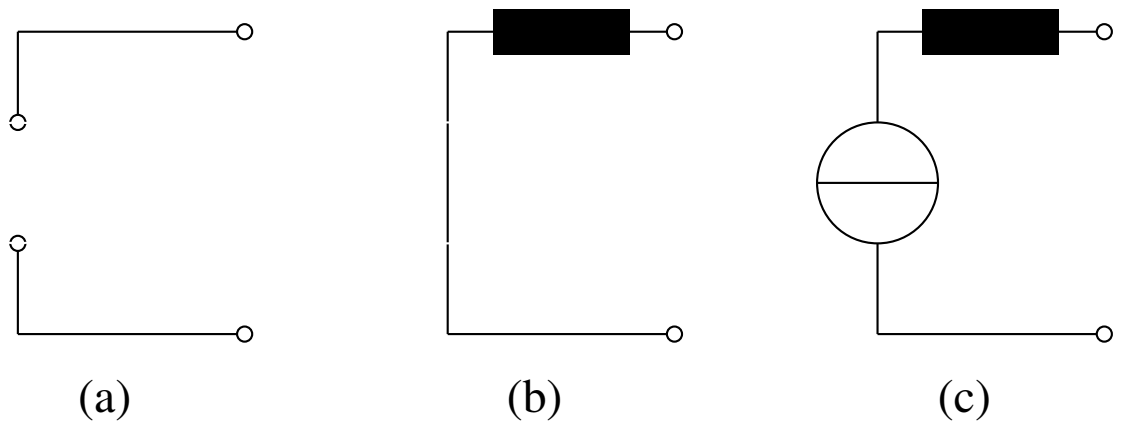
	The basis of the entire Classical Reactance Method is the absence of sources in the negative sequence on both sides, B\textsubscript{1} and B\textsubscript{2}.
	If the local source is a GFL IBR this assumption is no longer correct.
	These converters are represented by a current source in symmetrical components, both in the positive and negative sequence.
	The behavior of the current source in the negative sequence depends on the FRT requirements of the respective grid code.
	Two types are common.
	First, a pure current injection in the positive sequence, also during unbalanced grid situations.
	Consequently, the negative sequence current is controlled to zero.
	This is shown in Fig.~\ref{fig:2_GFL IBR NS_a}.		
	Secondly, a feed of reactive current in the negative sequence.
	Inductive current is fed to reduce the negative sequence voltage and thus balance the grid voltage.
	Mostly this is done via a droop control.
	The higher the negative sequence voltage, the more inductive current is fed.
	This imitates the behavior of a reactance, shown in  Fig.~\ref{fig:2_GFL IBR NS_b}.
	Still, due to current limitations of IBRs, an overall reactance imitating behavior is not possible.
	This is shown in Fig.~\ref{fig:2_GFL IBR NS_c}, as a current source and impedance.
	While there were no requirements in the past, negative sequence injection during unbalanced grid faults is now required in an increasing number of grid codes.
	Still, many existing GFL IBRs do not follow these new requirements, as they were commissioned according to older grid codes.
	In addition, negative sequence current injection is partially unreliable\cite{Nagpal2025}.
	If negative sequence injection is suppressed, the Classical Reactance Method algorithm in \eqref{eq:2_Loop_Equation_Imag_4} cannot be solved adequately.
	For this case a new fault current substitution is developed and introduced in th following chapter.
	
	\section{New Fault Current Substitution}
	\label{C:3_New Fault Current Substitution}
	
	As the negative sequence source impedance of the converter varies with the grid code and control mode, it may be unknown.
	Therefore, the algorithm must be changed to be independent of this variable and thus the negative sequence current.
	\begin{figure}[!t]
		\centering
		\begin{tikzpicture}[scale=0.57]
			\draw (0,-5) -- (12,-5);
			\draw (0,-8) -- (12,-8);
			\draw (0,-5) -- (0,-6.1);
			\draw (0,-6.9) -- (0,-8);
			\draw (12,-5) -- (12, -8);
			\draw (6,-5) -- (6,-4);
			\draw (6,-8) -- (6,-9);
			\draw[draw=black, dash pattern=on 2pt off 1pt, fill=white] (-0.5,-4.8) rectangle (1.5,-8.2);
			\node at (0.5, -6.5) {\huge{?}};
			\fill[black] (10.5,-4.825) rectangle (11.5,-5.175);
			\node[scale=0.9, transform shape] at (3, -4.75) {$\mathbf{B_1^-}$};
			\draw[fill=white, draw=black] (3,-5) circle [radius=0.05];
			\draw[fill=white, draw=black] (3,-8) circle [radius=0.05];
			\node[scale=0.9, transform shape] at (9, -4.75) {$\mathbf{B_2^-}$};
			\draw[fill=white, draw=black] (9,-5) circle [radius=0.05];
			\draw[fill=white, draw=black] (9,-8) circle [radius=0.05];
			\fill[black] (4,-4.825) rectangle (5,-5.175);
			\fill[black] (7,-4.825) rectangle (8,-5.175);
			\node at (11, -4.55) {\scriptsize$\underline{Z}_{\mathrm{B2}}$};
			\node at (4.5, -4.55) {\scriptsize$m\underline{Z}_{\mathrm{L}}$};
			\node at (7.5, -4.55) {\scriptsize$(1-m)\underline{Z}_{\mathrm{L}}$};
			\draw[->, stealth-] (6,-4.4 -- ++(0, -0.01);
			\node at (5.6, -4.4) {\scriptsize$\underline{I}^-_{\mathrm{F}}$};
			\draw[->, stealth-] (2.25,-5) -- ++(-0.01, 0);
			\node at (2.25, -4.6) {\scriptsize$\underline{I}^-_{\mathrm{B1}}$};
			\draw[->, stealth-] (9.75,-5) -- ++(0.01, 0);
			\node at (9.75, -4.6) {\scriptsize$\underline{I}^-_{\mathrm{B2}}$};
			\draw[->, stealth-] (6,-7.8) -- ++(0, 2.6);
			\node at (6.5, -6.5) {\scriptsize$\underline{V}^-_{\mathrm{c}}$};
			\draw[->, stealth-] (3,-7.8) -- ++(0, 2.6);
			\node at (3.6, -6.5) {\scriptsize$\underline{V}^-_{\mathrm{B1}}$};
			\draw[->, stealth-] (4.9,-5.3) -- ++(-0.8, 0);
			\node at (4.5, -5.65) {\scriptsize$\underline{V}^-_{\mathrm{mZL}}$};
			\draw[->, stealth-] (7.1,-5.3) -- ++(0.8, 0);
			\node at (7.5, -5.65) {\scriptsize$\underline{V}^-_{\mathrm{(1-m)ZL}}$};
			\draw[->, stealth-] (10.6,-5.3) -- ++(0.8, 0);
			\node at (11, -5.65) {\scriptsize$\underline{V}^-_{\mathrm{Z,B2}}$};
		\end{tikzpicture}
		\caption{Equivalent circuit of negative sequence during 2pF with fault resistance and remote feed}
		\label{fig:3_2pF in SymC_NS}
	\end{figure}
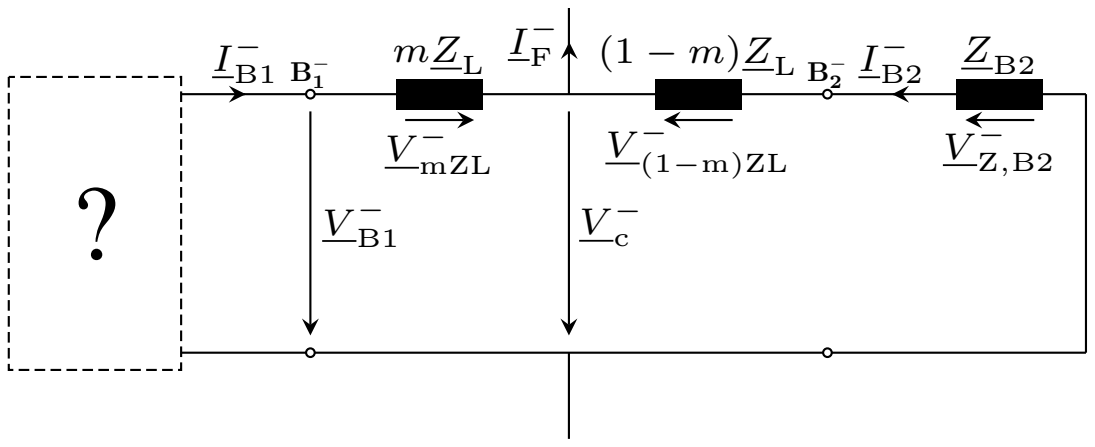
	
	Fig.~\ref{fig:3_2pF in SymC_NS} shows the negative sequence during the 2pF.
	The source type and impedance of the GFL IBR are undefined.
	Still, the negative sequence fault current can be expressed as in \eqref{eq:3_IF_substitution}, assuming that the external grid behaves as a short-circuit behind an impedance in the negative sequence.
	\begin{equation}
		\underline{I}^-_\mathrm{F} = \underline{I}^-_\mathrm{B1} + \underline{I}^-_\mathrm{B2}
		\label{eq:3_IF_substitution}
	\end{equation}
	While the local current is measured, the remote current must be calculated.
	For this reason the coupling voltage $\underline{V}^-_\mathrm{c}$ is introduced.
	In \eqref{eq:3_I_R_substitution} the remote negative sequence current is expressed depending on this voltage.
	\begin{equation}
		\underline{I}^-_\mathrm{B2} = \frac{-\underline{V}^-_\mathrm{c}}{\underline{Z}_\mathrm{B2}+(1-m)\underline{Z}_\mathrm{L}}
		\label{eq:3_I_R_substitution}
	\end{equation}
	As in the Classical Reactance Method it is assumed that the source impedance is known and the fault location is estimated to the zone boundary of the Distance Protection.
	The coupling voltage itself can be expressed by \eqref{eq:3_V_c_substitution}.
	\begin{equation}
		\underline{V}^-_\mathrm{c} = \underline{V}^-_\mathrm{B1} - \underline{V}^-_\mathrm{mZL} = \underline{V}^-_\mathrm{B1} - \underline{I}^-_\mathrm{B1}m\underline{Z}_\mathrm{L}
		\label{eq:3_V_c_substitution}
	\end{equation}
	Therefore, KVL is applied on the left side of the equivalent circuit in Fig. \ref{fig:3_2pF in SymC_NS}.
	With \eqref{eq:3_I_R_substitution} and \eqref{eq:3_V_c_substitution}, \eqref{eq:3_IF_substitution} is rewritten as \eqref{eq:3_IF_substitution_new}.
	\begin{equation}
		\begin{split}
			\underline{I}^-_\mathrm{F} &= \underline{I}^-_\mathrm{B1} + \frac{-\underline{V}^-_\mathrm{B1} + \underline{I}^-_\mathrm{B1}m\underline{Z}_\mathrm{L}}{\underline{Z}_\mathrm{B2}+(1-m)\underline{Z}_\mathrm{L}}
			\\
			&= \frac{-\underline{V}^-_\mathrm{B1} + \underline{I}^-_\mathrm{B1} \left( \underline{Z}_\mathrm{L} + \underline{Z}_\mathrm{B2} \right) }{\underline{Z}_\mathrm{B2}+(1-m)\underline{Z}_\mathrm{L}}
		\end{split}
		\label{eq:3_IF_substitution_new}
	\end{equation}
	The fault current is therefore substituted by only known or measurable values.
	With \eqref{eq:3_IF_substitution_new}, the original calculation for the fault current in \eqref{eq:2_Fault current in abc} is rewritten to \eqref{eq:3_I_substitution_new}.
	\begin{equation}
		\begin{split}
			\underline{I}_\mathrm{F} = \left(\underline{a}-\underline{a}^2 \right) \frac{-\underline{V}^-_\mathrm{B1} + \underline{I}^-_\mathrm{B1} \left( \underline{Z}_\mathrm{L} + \underline{Z}_\mathrm{B2} \right) }{\underline{Z}_\mathrm{B2}+(1-m)\underline{Z}_\mathrm{L}} = \underline{I}^{-*}_\mathrm{subst}
		\end{split}
		\label{eq:3_I_substitution_new}
	\end{equation}
	This fault current is defined as the new substituted negative sequence current $\underline{I}^{-*}_\mathrm{subst}$.
	Using the same procedure as in the previous chapter for the Classical Reactance Method, the line reactance up to the fault can now be measured without any knowledge of the internal negative sequence source, shown in \eqref{eq:3_X_meas_new}.
	\begin{equation}
		\begin{split}
			X_\mathrm{meas} &= X_\mathrm{L} = \frac{\Im \left\{ \left( \underline{V}_\mathrm{b}-\underline{V}_\mathrm{c}\right)  \overline{\underline{I}}^{-*}_\mathrm{subst} \right\}}{\Im \left\{ e^{j\varphi_{\mathrm{\underline{Z}L}}} \left( \underline{I}_\mathrm{b}-\underline{I}_\mathrm{c}\right)  \overline{\underline{I}}^{-*}_\mathrm{subst} \right\}} \sin \left( \varphi_{\mathrm{\underline{Z}L}} \right)
		\end{split}
		\label{eq:3_X_meas_new}
	\end{equation}
	This novel algorithm is called the Advanced Reactance Method in the following.
	The algorithm is independent of the negative sequence converter control and thus also works without negative sequence injection by the GFL IBR.
	It can therefore be used either as a fallback for the reactance method if the measured negative sequence current is too small, or to replace the entire classical algorithm.
	
	Table~\ref{tab:Comparison} summarizes the measured variables and required parameters for Distance Protection, the Classical Reactance Method, and the Advanced Reactance Method.
	While all three methods rely on the measured phase-to-ground voltages and currents at the local bus, the Classical Reactance Method requires the calculation of the negative sequence current, and the Advanced Reactance Method additionally requires the negative sequence voltage.
	Distance Protection requires only the line impedance, whereas the Classical Reactance Method needs the estimated fault location and the negative sequence source impedances at both the local and remote buses.
	The Advanced Reactance Method requires the estimated fault location and the remote source impedance, but does not depend on the local negative sequence source impedance.
	
	\begin{table}[!b]
		\centering
		\caption{Requirements of Distance Protection, Classical Reactance Method and Advanced Reactance Method}
		\label{tab:Comparison}
		\renewcommand{\arraystretch}{1.3}
		\begin{tabular}{l l l}
			\toprule
			\bfseries Algorithm & \bfseries Measured Values & \bfseries Needed Parameters\\
			\midrule
			DP & $\underline{V}_\mathrm{B1}^{abc}$, $\underline{I}_\mathrm{B1}^{abc}$ & $\underline{Z}_\mathrm{L}$\\
			CRM & $\underline{V}_\mathrm{B1}^{abc}$, $\underline{I}_\mathrm{B1}^{abc}$, $\underline{I}_\mathrm{B1}^-$ & $\underline{Z}_\mathrm{L}$, $m$ (estimated), $\underline{Z}_\mathrm{B1}^-$,  $\underline{Z}_\mathrm{B2}^-$\\
			ARM & $\underline{V}_\mathrm{B1}^{abc}$, $\underline{I}_\mathrm{B1}^{abc}$, $\underline{I}_\mathrm{B1}^-$, $\underline{V}_\mathrm{B1}^-$ & $\underline{Z}_\mathrm{L}$, $m$ (estimated),  $\underline{Z}_\mathrm{B2}^-$\\
			
			\bottomrule
		\end{tabular}
	\end{table}
	
	\section{Parameterization of the Classical Reactance Method in the Presence of GFL IBR}
	\label{C:31_Parameterization of the Classical Reactance Method in the Presence of GFL IBR}
	The Advanced Reactance Method is not available for all protection relays.
	Therefore, the Classical Reactance Method can be parameterized according to GFL IBRs.
	The algorithm requires known source impedances of the local and remote side in the negative sequence.
	The GFL IBR is a voltage source connected across impedances such as converter impedances, filters or transformers.
	Still, these impedances do not determine the impedance at the Point of Common Coupling (PCC) in steady state, but the GFL IBR control, that imitates a current source behavior.
	The control must therefore be accurately represented in terms of impedance so that the Classical Reactance Method remains reliable.
	Typical grid codes demand an inductive effect in the negative sequence to decrease the negative sequence voltage during unsymmetrical faults.
	As no active current is required or reasonable the angle of the impedance is set to inductive reactive, $90^\circ$.
	The impedance value can be calculated using the power rating of the GFL IBR and the rated voltage.
	However, it is important to note that even in the case of an unbalanced fault, not all of the current is fed in the negative sequence.
	The proportion of current in the positive and negative sequences depends on current limitation and prioritization.
	This is not uniformly defined.
	However, it can be assumed that no more than half of the current is fed into the negative sequence.
	Lastly, the overcurrent factor during FRT must be taken into account.
	GFL IBRs typically feed $1.1$ to $\SI{1.5}{pu}$ short-circuit current.
	Based on these assumptions, the impedance of the GFL IBR in the negative sequence is given by \eqref{eq:3_GFL IBR NS Z}.
	The parameterization is evaluated in \ref{C:4_Application and Parameterization of the Reactance Method for GFL IBRs}.
	\begin{equation}
		\underline{Z}_\mathrm{IBR}^- = jX_\mathrm{IBR}^{-*} = 2\cdot \frac{V_\mathrm{n}^2}{f_\mathrm{FRT}\cdot S_\mathrm{IBR}}\cdot e^{j90^\circ}
		\label{eq:3_GFL IBR NS Z}
	\end{equation}
	
	\section{Evaluation}
	\label{C:4_Application and Parameterization of the Reactance Method for GFL IBRs}
	
	In the following, the findings of the previous chapters are evaluated.
	Therefore, the test grid is presented.
	First, the impact of different behaviors of the GFL IBR FRT on the measured reactance of the Distance Protection, Classical Reactance Method and Advanced Reactance Method is evaluated.
	Second, a realistic scenario is simulated.
	The GFL IBR follows a FRT with reactive current injection in the positive and negative sequence.
	Moreover, an external grid is added to the bus B\textsubscript{1}, leading to three possible configurations: Only external grid, external grid and GFL IBR, only GFL IBR.
	The robustness of the Distance Protection, Classical Reactance Method and Advanced Reactance Method to grid configuration changes is evaluated for two different fault resistances.
	Third, the parameterization of the Classical Reactance Method in the presence of GFL IBRs is evaluated.
	
	\subsection{Test Grid for Evaluation}
	\label{C:41_Test Grid for Evaluation}
	\begin{figure}[!t]
		\centering
		\begin{tikzpicture}[scale=0.7]
			\draw (1.6,-0.4) rectangle (2.4,0.4);
			\draw (1.6,-0.4) -- (2.4,0.4);				
			\draw (2.0,-0.15) sin (2.075,-0.05) cos (2.15,-0.15) sin (2.225,-0.25) cos (2.30,-0.15);
			\draw (1.7, 0.2) -- (2.0, 0.2);
			\draw (1.7, 0.1) -- (2.0, 0.1);
			\node[align=center, inner sep=0pt] at (2,-1) {GFL\\[-2pt]IBR};
			\node[align=center, inner sep=0pt] at (11.6+0.65,-1.25) {External\\[-2pt]Grid};
			\draw (11.6,-0.65) rectangle (12.9,0.65);
			\draw (11.6,-0.45) -- (12.7,0.65);
			\draw (11.6,-0.25) -- (12.5,0.65);
			\draw (11.6,-0.05) -- (12.3,0.65);
			\draw (11.6, 0.15) -- (12.1,0.65);
			\draw (11.6, 0.35) -- (11.9,0.65);
			\draw (11.6,-0.65) -- (12.9,0.65);
			\draw (11.8,-0.65) -- (12.9,0.45);
			\draw (12.0,-0.65) -- (12.9,0.25);
			\draw (12.2,-0.65) -- (12.9,0.05);
			\draw (12.4,-0.65) -- (12.9,-0.15);
			\draw (12.6,-0.65) -- (12.9,-0.35);
			\draw (11.6,0.45) -- (12.7,-0.65);
			\draw (11.6,0.25) -- (12.5,-0.65);
			\draw (11.6,0.05) -- (12.3,-0.65);
			\draw (11.6,-0.15) -- (12.1,-0.65);
			\draw (11.6,-0.35) -- (11.9,-0.65);
			\draw (11.6,0.65) -- (12.9,-0.65);
			\draw (11.8,0.65) -- (12.9,-0.45);
			\draw (12.0,0.65) -- (12.9,-0.25);
			\draw (12.2,0.65) -- (12.9,-0.05);
			\draw (12.4,0.65) -- (12.9,0.15);
			\draw (12.6,0.65) -- (12.9,0.35);
			\draw (12.8,-0.65) -- (12.9,-0.55);
			\draw (12.8,0.65) -- (12.9,0.55);
			\draw (11.6,0.55) -- (11.7,0.65);
			\draw (11.6,-0.55) -- (11.7,-0.65);
			\draw (3, -0.5) -- (3, 0.5);
			\node at (3, 0.75) {$\mathbf{B_1}$};
			\draw (3, -0.5) -- (3, 0.5);
			\node at (11, 0.75) {$\mathbf{B_2}$};
			\draw (11, -0.5) -- (11, 0.5);
			\draw (2.4, 0) -- (11.6, 0);
			\node at (8.25,0.41) {\Huge\Lightning};
			\draw (8, 0) -- (8, -1);
			\draw[fill=white, draw=black] (7.9,-0.2) rectangle (8.1,-0.8);
			\node at (8.6, -0.52) {$R_{\mathrm{F}}$};
			\draw[<->, stealth-stealth] (3,-1.6) -- ++ (5, 0);
			\draw[<->, stealth-stealth] (8,-1.6) -- ++ (3, 0);
			\node at (5.5, -1.25) {$m\underline{Z}_{\mathrm{L}}$};
			\node at (9.5, -1.25) {$(1-m)\underline{Z}_{\mathrm{L}}$};
			\draw[fill=black, draw=black] (3.3,-0.1) rectangle (3.5,0.1);
			\node at (3.4, -0.4) {$\mathrm{R}_{\mathbf{1}}$};
			\draw[red] (3.4,-0.25) ellipse (0.35 and 0.6);
			\draw[fill=black, draw=black] (10.7,-0.1) rectangle (10.5,0.1);
			\node at (10.6, -0.4) {$\mathrm{R}_{\mathbf{2}}$};
		\end{tikzpicture}
		\caption{Test setup for evaluation of robustness of Distance Protection, Classical Reactance Method and Advanced Reactance Method against changing GFL IBR FRT behavior and parameterization of Classical Reactance Method}
		\label{fig:4_Test Grid}
	\end{figure}
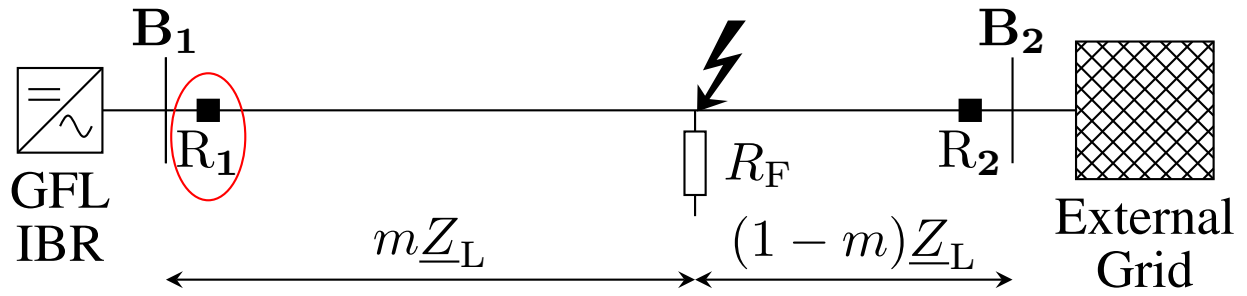
	\begin{table}[!b]
		\centering
		\caption{Grid and GFL IBR parameters of the test grid}
		\label{tab:Evaluation}
		\renewcommand{\arraystretch}{1.3}
		\begin{tabular}{llp{0.3\textwidth}}
			\toprule
			\bfseries Variable & \bfseries Value & \bfseries Description \\
			\midrule
			$V_\mathrm{n}$ & $\SI{380}{kV}$ & Grid Voltage \\
			$f_\mathrm{n}$ & $\SI{50}{Hz}$ & Grid frequency \\
			$S_\mathrm{k,G}^{''}$ & $\SI{10}{GVA}$ & Short-circuit power of external grid\\
			$\left( \frac{R}{X}\right)_\mathrm{G}$ & $\SI{0.1}{}$ & R/X ratio of external grid\\
			$l_\mathrm{L}$ & $\SI{50}{km}$ & Length of transmission line \\
			$R_\mathrm{L}^{'}$ & $\SI{0.009}{\ohm\per\kilo\meter}$ & Specific resistance of transmission line \\
			$X_\mathrm{L}^{'}$ & $\SI{0.242}{\ohm\per\kilo\meter}$ & Specific reactance of transmission line \\
			$S_\mathrm{C}$ & $\SI{1}{GVA}$ & Rated power of GFL IBR\\
			$S_\mathrm{op,C}$ & $\SI{750}{MW}$ & Pre-fault power operation of GFL IBR \\
			$R_\mathrm{C}$ & $\SI{0.5474}{\Omega}$ & Resistance of GFL IBR and Transformer \\
			$X_\mathrm{C}$ & $\SI{5.6512}{\Omega}$ & Reactance of GFL IBR and Transformer \\
			$f_\mathrm{FRT}$ & $\SI{1.2}{}$ & Overcurrent factor of GFL IBR during FRT \\
			$K_\mathrm{IrV}^{+}$ & $\SI{4}{}$ & Droop factor of reactive current during FRT of GFL IBR in the positive sequence \\
			$K_\mathrm{IrV}^{-}$ & $\SI{2}{}$ & Droop factor of reactive current during FRT of GFL IBR in the negative sequence\\
			\bottomrule
		\end{tabular}
	\end{table}
	Fig.~\ref{fig:4_Test Grid} shows the test grid.
	A GFL IBR at bus B\textsubscript{1} is connected to an external grid at bus B\textsubscript{2} via a transmission line.
	All relevant grid and IBR parameters are listed in Table~\ref{tab:Evaluation}.
	The protection relay R\textsubscript{1} at bus B\textsubscript{1} is examined.
	The GFL IBR control consists of a positive and negative sequence current control loop.
	Starting with a Double Decoupled Synchronous Reference Frame (DDSRF) and a Phase-Locked-Loop (PLL) to synchronize with the grid angle and separate the measured currents and voltages at the PCC (B\textsubscript{1}) into positive and negative sequence dq-values.
	The high-level current control then follows typical FRT in grid codes like \cite{IEEE_standard} or \cite{VDE-AR-N-4130-2018}, which mandate reactive current injection during unbalanced faults to support grid voltage.
	In the positive sequence, a reactive current-to-voltage droop is implemented.
	This means, that a increasing capacitive reactive current is injected for a decreasing voltage.
	This increases the grid voltage during the fault.
	In the negative sequence the same mechanism is implemented.
	Therefore, an inductive current is injected when the negative sequence voltage increases due to a grid imbalance.
	This leads to a decrease in the negative sequence grid voltage and therefore balancing of the grid voltage.
	As the maximum current of an IBR is limited, the set currents of the high-level control must be limited.
	An unbalanced current limitation is implemented, prioritizing the reactive currents in the positive and negative sequence over the active current in the positive sequence\cite{Teodorescu_2011,Lee_2011,Winjnhove_2014}.
	The reactive currents in the positive and negative sequence are prioritized equally.
	This corresponds to a typical FRT.
	The high-level control apart from the FRT will not be considered further, as it is not relevant to this evaluation.
	It is possible to change the FRT to a pure active current injection in the positive sequence, a pure capacitive reactive current injection in the positive sequence, a pure inductive reactive current injection in the negative sequence, or combination of those.
	The GFL IBR control is based on \cite{Teodorescu_2011}.
	
	\subsection{Evaluation of different GFL IBR Fault-Ride-Through}
	\label{C:43_Evaluation of different GFL IBR Fault-Ride-Trough}
\begin{figure}[!t]
		\centering
		\captionsetup[subfigure]{skip=0pt}
		\setlength{\abovecaptionskip}{2pt}
		\setlength{\belowcaptionskip}{-5pt}
		
		\begin{subfigure}[t]{\columnwidth}
			\centering
			\begin{tikzpicture}
				\begin{axis}[
					width=\linewidth,
					height=4.45cm,
					xlabel={$m$ in \%},
					xlabel={},
					ylabel={Deviation in \%},
					ylabel style={yshift=-10pt},
					xtick={10,20,30,40,50,60,70,80,90},
					legend style={
						font=\footnotesize,
						at={(0.5,1.225)},
						anchor=north,
						legend columns=4,
						draw=none
					},
					grid=major,
					grid style={line width=0.3pt, draw=gray!30},
					xmin=0, xmax=100,
					ymin=-5, ymax=15,
					]
					\addplot[color=green!70!black, mark=*, mark size=1.5pt, line width=0.6pt]
					coordinates {
						(10,5.8224) (20,6.263) (30,6.7622) (40,7.3315) (50,7.9855)
						(60,8.7429) (70,9.6279) (80,10.672) (90,11.919)
					};
					\addlegendentry{DP}
					\addplot[color=blue, mark=triangle*, mark size=2.5pt, line width=0.6pt]
					coordinates {
						(10,-0.7316) (20,-0.6804) (30,-0.6173) (40,-0.5392) (50,-0.4426)
						(60,-0.3212) (70,-0.169) (80,0.024) (90,0.2712)
					};
					\addlegendentry{CRM}
					\addplot[color=red, mark=x, mark size=2.5pt, line width=0.6pt]
					coordinates {
						(10,-0.7018) (20,-0.6456) (30,-0.577) (40,-0.4928) (50,-0.389)
						(60,-0.2603) (70,-0.0993) (80,0.1037) (90,0.3624)
					};
					\addlegendentry{ARM}
					\addplot[black, thin, dashed] coordinates {(0,0) (100,0)};
					\addplot[black, thin, dashed] coordinates {(80,-1000) (80,100)};
				\end{axis}
				\node[anchor=north, yshift=-16pt] at (current axis.south) {(a)};
			\end{tikzpicture}
			\captionsetup{labelformat=empty}
			\caption{}
			\label{fig:Eval_IBR_Q}
		\end{subfigure}
		\hfill
		\begin{subfigure}[t]{\columnwidth}
			\centering
			\begin{tikzpicture}
				\begin{axis}[
					width=\linewidth,
					height=4.45cm,
					xlabel={},
					ylabel={Deviation in \%},
					ylabel style={yshift=-10pt},
					xtick={10,20,30,40,50,60,70,80,90},
					legend style={
						font=\footnotesize,
						at={(0.5,1.225)},
						anchor=north,
						legend columns=4,
						draw=none
					},
					grid=major,
					grid style={line width=0.3pt, draw=gray!30},
					xmin=0, xmax=100,
					ymin=-65, ymax=3,
					]
					\addplot[color=green!70!black, mark=*, mark size=1.5pt, line width=0.6pt]
					coordinates {
						(10,-40.718) (20,-42.781) (30,-45.025) (40,-47.472) (50,-50.152)
						(60,-53.098) (70,-56.35) (80,-59.957) (90,-63.976)
					};
					\addlegendentry{DP}
					\addplot[color=blue, mark=triangle*, mark size=2.5pt, line width=0.6pt]
					coordinates {
						(10,-1.6318) (20,-1.55) (30,-1.4456) (40,-1.3126) (50,-1.143)
						(60,-0.9262) (70,-0.6482) (80,-0.29) (90,0.1745)
					};
					\addlegendentry{CRM}
					\addplot[color=red, mark=x, mark size=2.5pt, line width=0.6pt]
					coordinates {
						(10,-1.1804) (20,-1.0696) (30,-0.9332) (40,-0.7649) (50,-0.556)
						(60,-0.2953) (70,0.0319) (80,0.4458) (90,0.9734)
					};
					\addlegendentry{ARM}					
					\addplot[black, thin, dashed] coordinates {(0,0) (100,0)};
					\addplot[black, thin, dashed] coordinates {(80,-1000) (80,100)};
				\end{axis}
				\node[anchor=north, yshift=-16pt] at (current axis.south) {(b)};
			\end{tikzpicture}
			\captionsetup{labelformat=empty}
			\caption{}
			\label{fig:Eval_IBR_P}
		\end{subfigure}
		\hfill
		\begin{subfigure}[t]{\columnwidth}
			\centering
			\begin{tikzpicture}
				\begin{axis}[
					width=\linewidth,
					height=4.45cm,
					xlabel={},
					ylabel={Deviation in \%},
					ylabel style={yshift=-10pt},
					xtick={10,20,30,40,50,60,70,80,90},
					legend style={
						font=\footnotesize,
						at={(0.5,1.225)},
						anchor=north,
						legend columns=4,
						draw=none
					},
					grid=major,
					grid style={line width=0.3pt, draw=gray!30},
					xmin=0, xmax=100,
					ymin=-2, ymax=17,
					]
					\addplot[color=green!70!black, mark=*, mark size=1.5pt, line width=0.6pt]
					coordinates {
						(10,7.0156) (20,7.562) (30,8.1819) (40,8.8896) (50,9.7035)
						(60,10.647) (70,11.7507) (80,13.0546) (90,14.6123)
					};
					\addlegendentry{DP}
					\addplot[color=red, mark=x, mark size=2.5pt, line width=0.6pt]
					coordinates {
						(10,-0.7092) (20,-0.6525) (30,-0.5832) (40,-0.498) (50,-0.3928)
						(60,-0.2619) (70,-0.098) (80,0.1093) (90,0.374)
					};
					\addlegendentry{ARM}
					\addplot[black, thin, dashed] coordinates {(0,0) (100,0)};
					\addplot[black, thin, dashed] coordinates {(80,-1000) (80,100)};
				\end{axis}
				\node[anchor=north, yshift=-16pt] at (current axis.south) {(c)};
			\end{tikzpicture}
			\captionsetup{labelformat=empty}
			\caption{}
			\label{fig:Eval_IBR_Q_ONS}
		\end{subfigure}
		\hfill
		\begin{subfigure}[t]{\columnwidth}
			\centering
			\begin{tikzpicture}
				\begin{axis}[
					width=\linewidth,
					height=4.45cm,
					xlabel={$m$ in \%},
					ylabel={Deviation in \%},
					ylabel style={yshift=-10pt},
					xtick={10,20,30,40,50,60,70,80,90},
					legend style={
						font=\footnotesize,
						at={(0.5,1.225)},
						anchor=north,
						legend columns=4,
						draw=none
					},
					grid=major,
					grid style={line width=0.3pt, draw=gray!30},
					xmin=0, xmax=100,
					ymin=-75, ymax=10,
					]
					\addplot[color=green!70!black, mark=*, mark size=1.5pt, line width=0.6pt]
					coordinates {
						(10,-44.7055) (20,-46.7559) (30,-49.0074) (40,-51.4922) (50,-54.2496)
						(60,-57.3281) (70,-60.7885) (80,-64.7078) (90,-69.1847)
					};
					\addlegendentry{DP}
					\addplot[color=red, mark=x, mark size=2.5pt, line width=0.6pt]
					coordinates {
						(10,-3.7008) (20,-3.5851) (30,-3.3615) (40,-2.9781) (50,-2.3578)
						(60,-1.3864) (70,0.1061) (80,2.3703) (90,5.7689)
					};
					\addlegendentry{ARM}					
					\addplot[black, thin, dashed] coordinates {(0,0) (100,0)};
					\addplot[black, thin, dashed] coordinates {(80,-1000) (80,100)};
				\end{axis}
				\node[anchor=north, yshift=-29pt] at (current axis.south) {(d)};
			\end{tikzpicture}
			\captionsetup{labelformat=empty}
			\caption{}
			\label{fig:Eval_IBR_P_ONS}
		\end{subfigure}
		\caption{Comparison of Distance Protection (DP), Classical Reactance Method (CRM) and Advanced Reactance Method (ARM) during different FRT modes of the GFL IBR: (a) reactive current injection in positive and negative sequence, (b) active current injection in positive sequence and reactive current injection in negative sequence, (c) reactive current injection in positive sequence, (d) active current injection in positive sequence.}
		\label{fig:all_comparison}
	\end{figure}
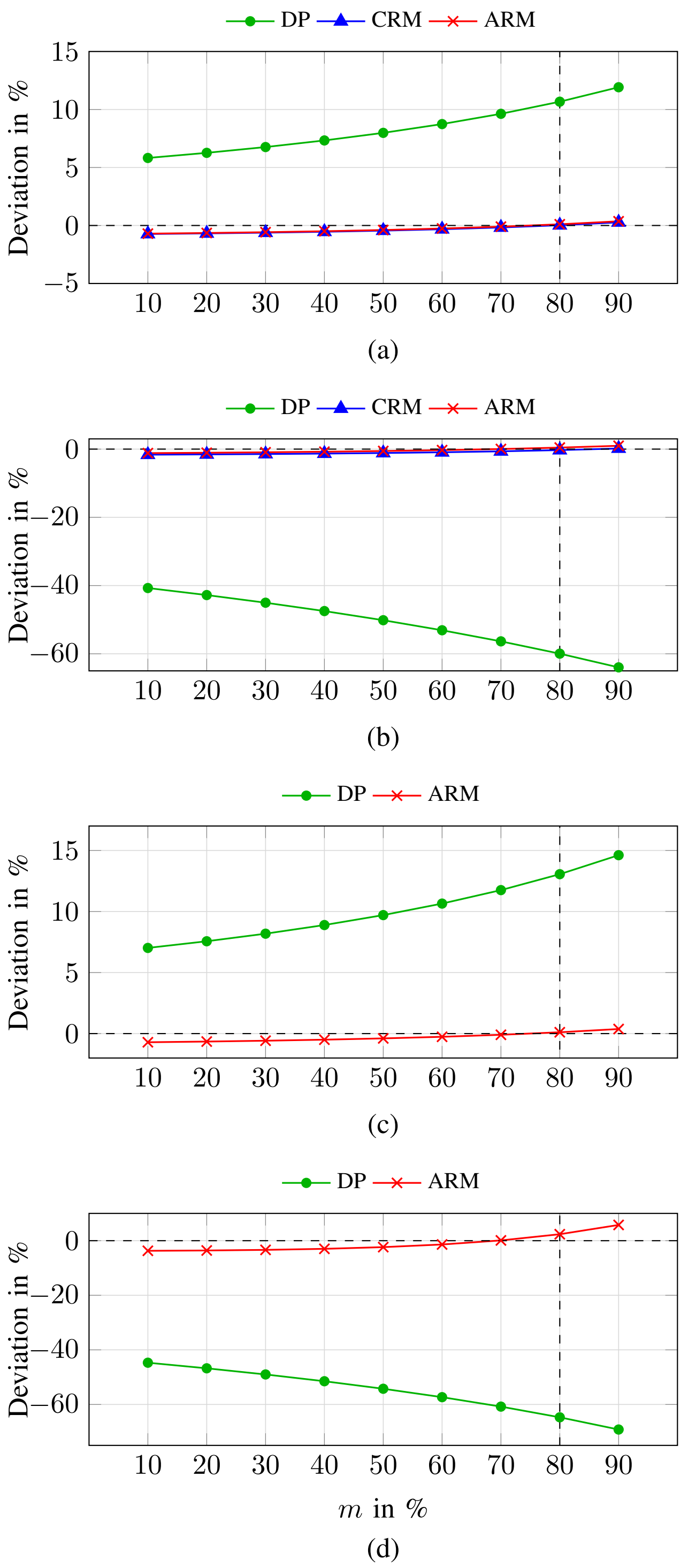

	In the following, the robustness of the Distance Protection, Classical Reactance Method and Advanced Reactance Method against varying behavior of the GFL IBR during FRT is evaluated.
	The fault resistance $R_\mathrm{F}$ is set to $\SI{2}{\Omega}$ while the fault location varies between 10\% and 90\%.
	The Classical Reactance Method is parameterized as proposed earlier.
	For the first scenario, the GFL IBR follows the typical FRT with reactive current injection in the positive and negative sequence.
	The deviation of the measured reactance to the actual reactance in percent of the line reactance for the Distance Protection, Classical Reactance Method and Advanced Reactance Method are shown in Fig.~\ref{fig:Eval_IBR_Q}.
	The Distance Protection shows a high deviation.
	The effect of an overestimated reactance during reactive power injection, as described in \cite{loebel_distance_nodate}, is evident.
	The Classical Reactance Method and Advanced Reactance Method show nearly identical results.
	Both show the typical behavior of a small underreach before the zone boundary and a small overreach after.
	Still the measurement deviations are little.
	At the most critical location, the zone boundary at 80\% of the line the Distance Protection has a measurement deviation of 10.6\%, while the Classical Reactance Method shows 0.024\% and the Advanced Reactance Method 0.1037\%.
	
	For the second scenario the GFL IBR injects active current in the positive sequence and reactive current in the negative sequence.
	The results for all three protection algorithms are shown in Fig.~\ref{fig:Eval_IBR_P}.
	The Distance Protection algorithm shows the opposite behavior than in the previous scenario.
	The deviation is negative, meaning an overreach of the protection relay, as described in \cite{loebel_distance_nodate}.
	Again, the Classical and Advanced Reactance Methods show very similar results.
	At the zone boundary, the Distance Protection has a measurement deviation of -60\%, while the Classical Reactance Method shows -0.29\% and the Advanced Reactance Method 0.45\%.
	
	For the third scenario, the GFL IBR only injects reactive current in the positive sequence during FRT, while the negative sequence current is suppressed.
	Consequently, the Classical Reactance Method cannot operate due to the lack of measured negative sequence current.
	The Distance Protection algorithm behaves similarly to the first scenario, but the deviation is even greater.
	The Advanced Reactance Method is almost unaffected and provides nearly the same results as in the first scenario.
	At the zone boundary, the Distance Protection shows a deviation of 13.05\%, while the Advanced Reactance Method shows 0.11\% deviation.
	
	For the last scenario, the GFL IBR only injects active current in the positive sequence during FRT, while the negative sequence current is suppressed again.
	Still, the Classical Reactance Method cannot operate due to the lack of measured negative sequence current.
	The Distance Protection algorithm behaves similarly to the second scenario, but the deviation is even greater again.
	This time the Advanced Reactance Method shows significantly larger measurement deviations, compared to the second scenario.
	Still, the deviation is much smaller than with Distance Protection.
	At the zone boundary, the Distance Protection shows a deviation of -64.71\%, while the Advanced Reactance Method shows 2.37\% deviation.
	
	Overall, the Distance Protection algorithm shows the worst robustness against changing GFL IBR FRT techniques.
	For all four cases, the deviation was the largest and it changed markedly due to active or reactive current injection.
	The correctly parameterized Classical Reactance Method shows good results for both active and reactive current injection in the positive sequence, as long as the negative sequence current is present.
	For the absence of negative sequence current the algorithm does not work anymore.
	The Advanced Reactance Method shows the best results, overall.
	Independent of the positive sequence current injection, the results are nearly the same as for the Classical Reactance Method.
	Without negative sequence current injection, the algorithm shows better results for reactive current injection in the positive sequence.
	
	\subsection{Evaluation of different Grid Topologies}
	\label{C:44_Classical Reactance Method and Distance Protection}
	
	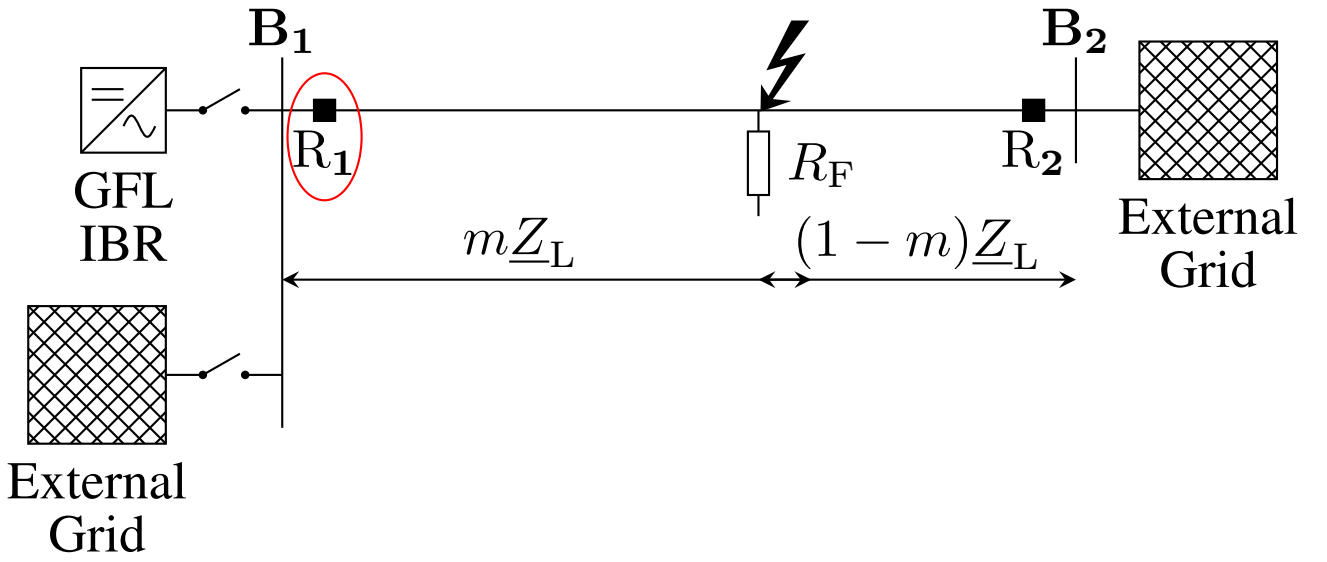
\begin{figure}[!t]
		\centering
		\begin{tikzpicture}[scale=0.7]
			\draw (1.6,-0.4) rectangle (2.4,0.4);
			\draw (1.6,-0.4) -- (2.4,0.4);				
			\draw (2.0,-0.15) sin (2.075,-0.05) cos (2.15,-0.15) sin (2.225,-0.25) cos (2.30,-0.15);
			\draw (1.7, 0.2) -- (2.0, 0.2);
			\draw (1.7, 0.1) -- (2.0, 0.1);
			\node[align=center, inner sep=0pt] at (2,-1) {GFL\\[-2pt]IBR};
			\draw (2.4, 0) -- (2.75, 0);
			\draw[fill=black, draw=black] (2.75,0) circle [radius=0.03];
			\draw (2.75, 0) -- (3.1, 0.2);
			\draw[fill=black, draw=black] (3.15,0) circle [radius=0.03];
			\draw (3.15, 0) -- (3.5, 0);
			\draw (2.4, -2.5) -- (2.75, -2.5);
			\draw[fill=black, draw=black] (2.75,-2.5) circle [radius=0.03];
			\draw (2.75, -2.5) -- (3.1, -2.3);
			\draw[fill=black, draw=black] (3.15,-2.5) circle [radius=0.03];
			\draw (3.15, -2.5) -- (3.5, -2.5);
			\node[align=center, inner sep=0pt] at (11.6+0.65,-1.25) {External\\[-2pt]Grid};
			\draw (11.6,-0.65) rectangle (12.9,0.65);
			\draw (11.6,-0.45) -- (12.7,0.65);
			\draw (11.6,-0.25) -- (12.5,0.65);
			\draw (11.6,-0.05) -- (12.3,0.65);
			\draw (11.6, 0.15) -- (12.1,0.65);
			\draw (11.6, 0.35) -- (11.9,0.65);
			\draw (11.6,-0.65) -- (12.9,0.65);
			\draw (11.8,-0.65) -- (12.9,0.45);
			\draw (12.0,-0.65) -- (12.9,0.25);
			\draw (12.2,-0.65) -- (12.9,0.05);
			\draw (12.4,-0.65) -- (12.9,-0.15);
			\draw (12.6,-0.65) -- (12.9,-0.35);
			\draw (11.6,0.45) -- (12.7,-0.65);
			\draw (11.6,0.25) -- (12.5,-0.65);
			\draw (11.6,0.05) -- (12.3,-0.65);
			\draw (11.6,-0.15) -- (12.1,-0.65);
			\draw (11.6,-0.35) -- (11.9,-0.65);
			\draw (11.6,0.65) -- (12.9,-0.65);
			\draw (11.8,0.65) -- (12.9,-0.45);
			\draw (12.0,0.65) -- (12.9,-0.25);
			\draw (12.2,0.65) -- (12.9,-0.05);
			\draw (12.4,0.65) -- (12.9,0.15);
			\draw (12.6,0.65) -- (12.9,0.35);
			\draw (12.8,-0.65) -- (12.9,-0.55);
			\draw (12.8,0.65) -- (12.9,0.55);
			\draw (11.6,0.55) -- (11.7,0.65);
			\draw (11.6,-0.55) -- (11.7,-0.65);
			\node at (3.5, 0.75) {$\mathbf{B_1}$};
			\draw (3.5, -3) -- (3.5, 0.5);
			\node at (11, 0.75) {$\mathbf{B_2}$};
			\draw (11, -0.5) -- (11, 0.5);
			\draw (3.5, 0) -- (11.6, 0);
			\node at (8.25,0.41) {\Huge\Lightning};
			\draw (8, 0) -- (8, -1);
			\draw[fill=white, draw=black] (7.9,-0.2) rectangle (8.1,-0.8);
			\node at (8.6, -0.52) {$R_{\mathrm{F}}$};
			\draw[<->, stealth-stealth] (3.5,-1.6) -- ++ (5, 0);
			\draw[<->, stealth-stealth] (8,-1.6) -- ++ (3, 0);
			\node at (5.75, -1.25) {$m\underline{Z}_{\mathrm{L}}$};
			\node at (9.5, -1.25) {$(1-m)\underline{Z}_{\mathrm{L}}$};
			\draw[fill=black, draw=black] (3.8,-0.1) rectangle (4,0.1);
			\node at (3.9, -0.4) {$\mathrm{R}_{\mathbf{1}}$};
			\draw[red] (3.9,-0.25) ellipse (0.35 and 0.6);
			\draw[fill=black, draw=black] (10.7,-0.1) rectangle (10.5,0.1);
			\node at (10.6, -0.4) {$\mathrm{R}_{\mathbf{2}}$};
			\draw (1.1,-3.15) rectangle (2.4,-1.85);
			\draw (1.1,-2.95) -- (2.2,-1.85);
			\draw (1.1,-2.75) -- (2.0,-1.85);
			\draw (1.1,-2.55) -- (1.8,-1.85);
			\draw (1.1,-2.35) -- (1.6,-1.85);
			\draw (1.1,-2.15) -- (1.4,-1.85);
			\draw (1.1,-3.15) -- (2.4,-1.85);
			\draw (1.3,-3.15) -- (2.4,-2.05);
			\draw (1.5,-3.15) -- (2.4,-2.25);
			\draw (1.7,-3.15) -- (2.4,-2.45);
			\draw (1.9,-3.15) -- (2.4,-2.65);
			\draw (2.1,-3.15) -- (2.4,-2.85);
			\draw (1.1,-2.05) -- (2.2,-3.15);
			\draw (1.1,-2.25) -- (2.0,-3.15);
			\draw (1.1,-2.45) -- (1.8,-3.15);
			\draw (1.1,-2.65) -- (1.6,-3.15);
			\draw (1.1,-2.85) -- (1.4,-3.15);
			\draw (1.1,-1.85) -- (2.4,-3.15);
			\draw (1.3,-1.85) -- (2.4,-2.95);
			\draw (1.5,-1.85) -- (2.4,-2.75);
			\draw (1.7,-1.85) -- (2.4,-2.55);
			\draw (1.9,-1.85) -- (2.4,-2.35);
			\draw (2.1,-1.85) -- (2.4,-2.15);
			\draw (2.3,-3.15) -- (2.4,-3.05);
			\draw (2.3,-1.85) -- (2.4,-1.95);
			\draw (1.1,-1.95) -- (1.2,-1.85);
			\draw (1.1,-3.05) -- (1.2,-3.15);
			\node[align=center, inner sep=0pt] at (1.1+0.65,-3.75) {External\\[-2pt]Grid};
		\end{tikzpicture}
		\caption{Test setup for evaluation of robustness of Distance Protection, Classical Reactance Method and Advanced Reactance Method against changing grid topologies.}
		\label{fig:4_Test Grid+EG}
	\end{figure}
	
	Now, the robustness of the Distance Protection, Classical Reactance Method and Advanced Reactance Method against changes in the grid topology is evaluated.
	Therefore, the GFL IBR FRT is set to typical grid code requirements.
	This consists of a positive and negative sequence reactive current injection.
	The prioritization of the current limitation is implemented as described before.
	The test grid, shown in Fig.~\ref{fig:4_Test Grid+EG}, is based on the previous test grid.
	This time, the feeding source at bus B\textsubscript{1} can be changed between a GFL IBR, an external grid or the combination.
	The external grid has the same parameters as the one at bus~B\textsubscript{2}.
	This scenario corresponds to typical topologies in the transmission grid.
	For example, buses, connected to large wind farms, HVDC links or grid boosters.
	If the transmission line to an external grid fails or is disconnected due to maintenance work, the protection at the substation still needs to work properly.
	All parameters of the test grid are the same as in the previous case, shown in Table.~\ref{tab:Evaluation}.
	Three scenarios are considered:
	
	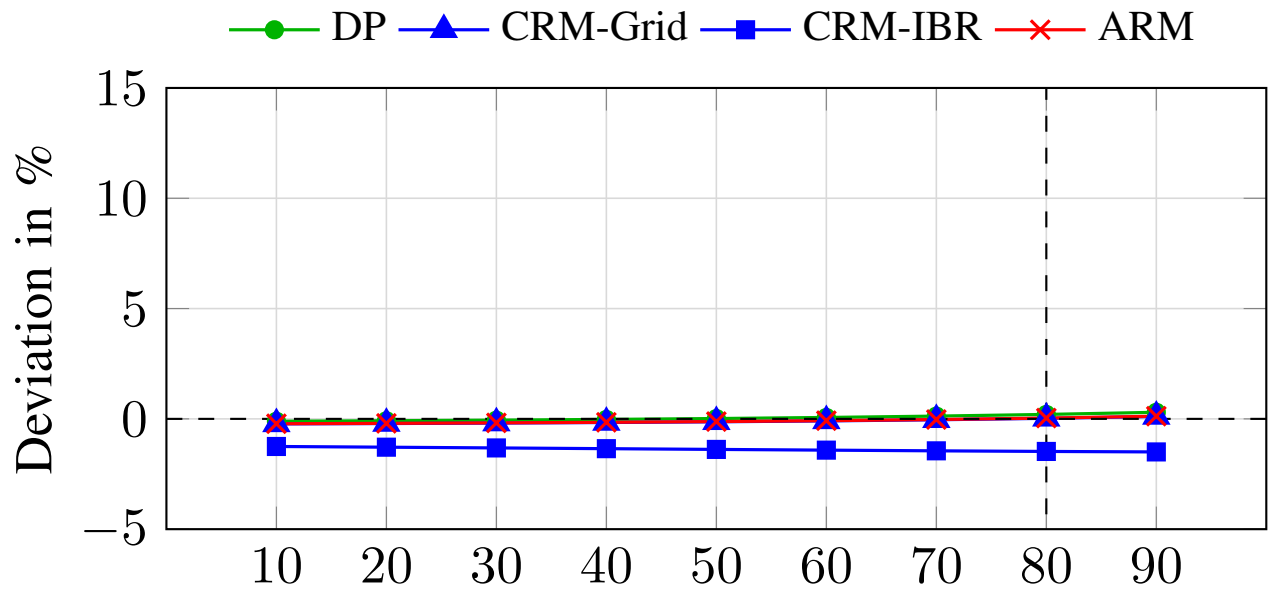
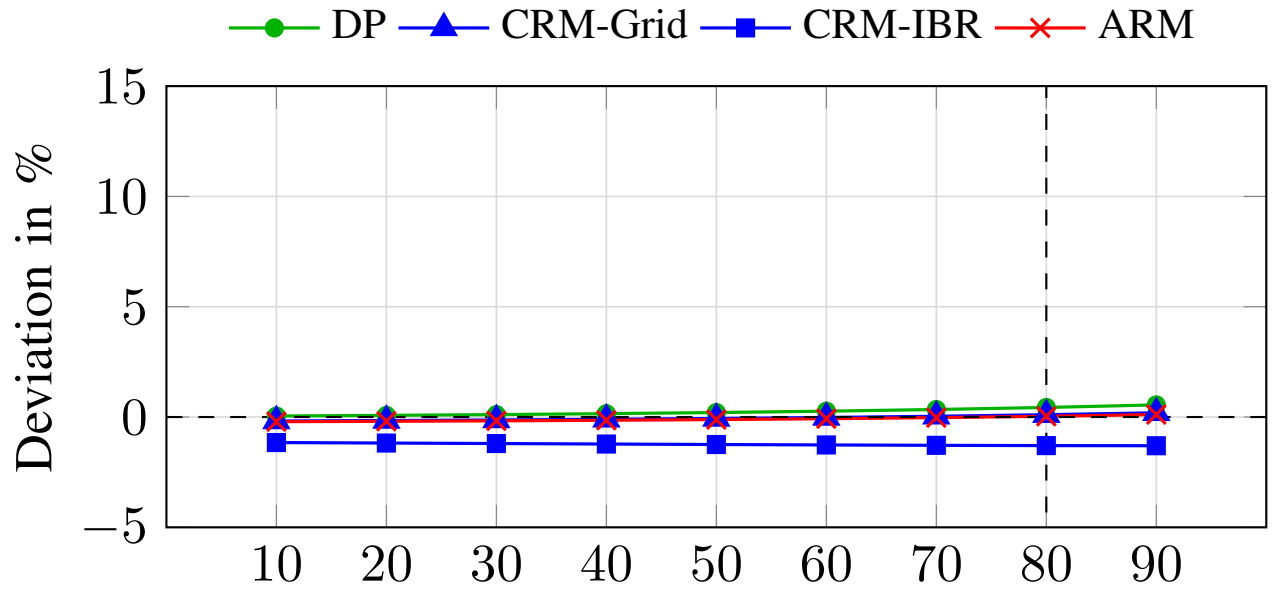
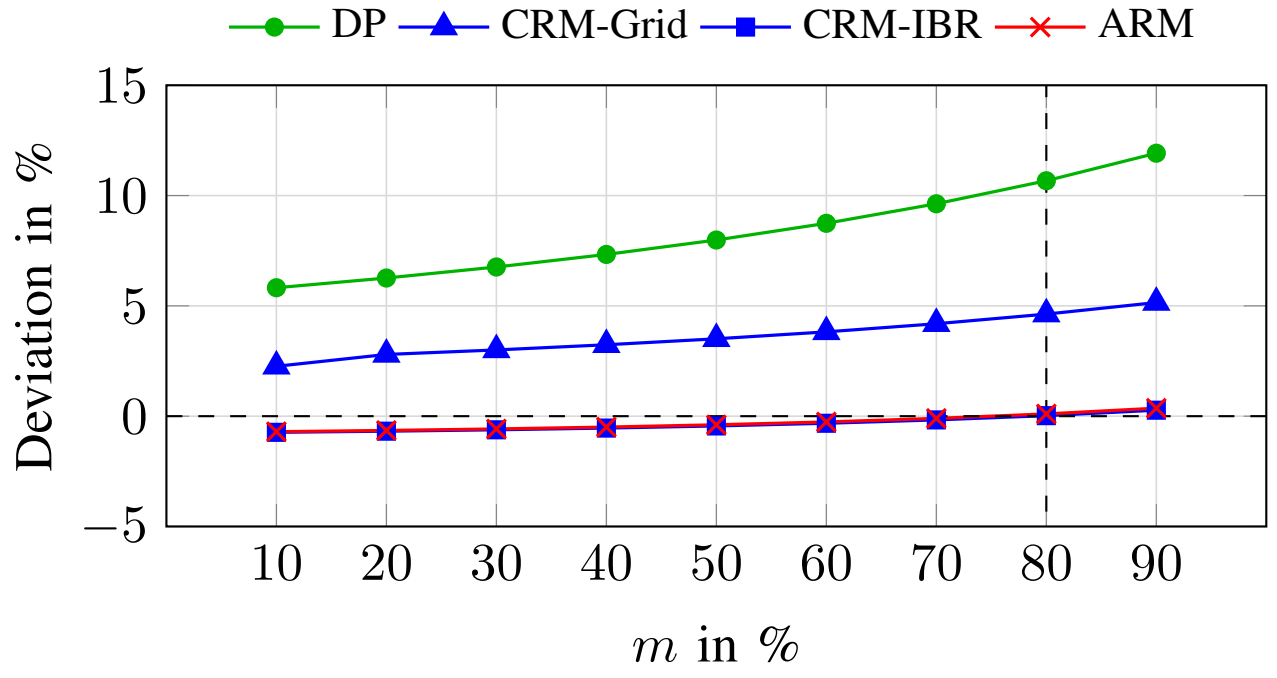
\begin{figure}[!t]
		\centering
		\captionsetup[subfigure]{skip=0pt}
		\setlength{\abovecaptionskip}{2pt}
		\setlength{\belowcaptionskip}{-5pt}
		
		\begin{subfigure}[t]{\columnwidth}
			\centering
			\begin{tikzpicture}
				\begin{axis}[
					width=\linewidth,
					height=4.5cm,
					xlabel={$m$ in \%},
					xlabel={},
					ylabel={Deviation in \%},
					ylabel style={yshift=-10pt},
					xtick={10,20,30,40,50,60,70,80,90},
					legend style={
						font=\footnotesize,
						at={(0.5,1.225)},
						anchor=north,
						legend columns=4,
						draw=none
					},
					grid=major,
					grid style={line width=0.3pt, draw=gray!30},
					xmin=0, xmax=100,
					ymin=-5, ymax=15,
					]
					\addplot[color=green!70!black, mark=*, mark size=1.5pt, line width=0.6pt]
					coordinates {
						(10,-0.1077) (20,-0.0850) (30,-0.0577) (40,-0.0247) (50,0.0154)
						(60,0.0647) (70,0.1256) (80,0.2019) (90,0.2985)
					};
					\addlegendentry{DP}
					\addplot[color=blue, mark=triangle*, mark size=2.5pt, line width=0.6pt]
					coordinates {
						(10,-0.2286) (20,-0.2117) (30,-0.1908) (40,-0.1648) (50,-0.1325)
						(60,-0.0919) (70,-0.0406) (80,0.0247) (90,0.1088)
					};
					\addlegendentry{CRM-Grid}
					\addplot[color=blue, mark=square*, mark size=1.5pt, line width=0.6pt]
					coordinates {
						(10,-1.2513) (20,-1.2837) (30,-1.317) (40,-1.3507) (50,-1.3844)
						(60,-1.4173) (70,-1.4485) (80,-1.4763) (90,-1.4981)
					};
					\addlegendentry{CRM-IBR}
					\addplot[color=red, mark=x, mark size=2.5pt, line width=0.6pt]
					coordinates {
						(10,-0.2207) (20,-0.2034) (30,-0.1821) (40,-0.1557) (50,-0.1228)
						(60,-0.0817) (70,-0.0298) (80,0.0362) (90,0.1211)
					};
					\addlegendentry{ARM}
					\addplot[black, thin, dashed] coordinates {(0,0) (100,0)};
					\addplot[black, thin, dashed] coordinates {(80,-1000) (80,100)};
				\end{axis}
				\node[anchor=north, yshift=-16pt] at (current axis.south) {(a)};
			\end{tikzpicture}
			\captionsetup{labelformat=empty}
			\caption{}
			\label{fig:EG_1}
		\end{subfigure}
		\hfill
		\begin{subfigure}[t]{\columnwidth}
			\centering
			\begin{tikzpicture}
				\begin{axis}[
					width=\linewidth,
					height=4.5cm,
					xlabel={},
					ylabel={Deviation in \%},
					ylabel style={yshift=-10pt},
					xtick={10,20,30,40,50,60,70,80,90},
					legend style={
						font=\footnotesize,
						at={(0.5,1.225)},
						anchor=north,
						legend columns=4,
						draw=none
					},
					grid=major,
					grid style={line width=0.3pt, draw=gray!30},
					xmin=0, xmax=100,
					ymin=-5, ymax=15,
					]
					\addplot[color=green!70!black, mark=*, mark size=1.5pt, line width=0.6pt]
					coordinates {
						(10,0.0464) (20,0.0752) (30,0.1102) (40,0.1526) (50,0.2038)
						(60,0.2658) (70,0.3414) (80,0.4344) (90,0.5502)
					};
					\addlegendentry{DP}
					\addplot[color=blue, mark=triangle*, mark size=2.5pt, line width=0.6pt]
					coordinates {
						(10,-0.1799) (20,-0.1586) (30,-0.1332) (40,-0.1027) (50,-0.0657)
						(60,-0.0204) (70,0.0354) (80,0.1051) (90,0.1929)
					};
					\addlegendentry{CRM-Grid}
					\addplot[color=blue, mark=square*, mark size=1.5pt, line width=0.6pt]
					coordinates {
						(10,-1.1566) (20,-1.1791) (30,-1.2017) (40,-1.224) (50,-1.2455)
						(60,-1.2655) (70,-1.2826) (80,-1.2954) (90,-1.3012)
					};
					\addlegendentry{CRM-IBR}
					\addplot[color=red, mark=x, mark size=2.5pt, line width=0.6pt]
					coordinates {
						(10,-0.2109) (20,-0.1937) (30,-0.1729) (40,-0.1473) (50,-0.1159)
						(60,-0.0769) (70,-0.0281) (80,0.0335) (90,0.1122)
					};
					\addlegendentry{ARM}
					\addplot[black, thin, dashed] coordinates {(0,0) (100,0)};
					\addplot[black, thin, dashed] coordinates {(80,-1000) (80,100)};
				\end{axis}
				\node[anchor=north, yshift=-16pt] at (current axis.south) {(b)};
			\end{tikzpicture}
			\captionsetup{labelformat=empty}
			\caption{}
			\label{fig:EG_2}
		\end{subfigure}
		\hfill
		\begin{subfigure}[t]{\columnwidth}
			\centering
			\begin{tikzpicture}
				\begin{axis}[
					width=\linewidth,
					height=4.5cm,
					xlabel={$m$ in \%},
					ylabel={Deviation in \%},
					ylabel style={yshift=-10pt},
					xtick={10,20,30,40,50,60,70,80,90},
					legend style={
						font=\footnotesize,
						at={(0.5,1.225)},
						anchor=north,
						legend columns=4,
						draw=none
					},
					grid=major,
					grid style={line width=0.3pt, draw=gray!30},
					xmin=0, xmax=100,
					ymin=-5, ymax=15,
					]
					\addplot[color=green!70!black, mark=*, mark size=1.5pt, line width=0.6pt]
					coordinates {
						(10,5.8224) (20,6.2630) (30,6.7622) (40,7.3315) (50,7.9855)
						(60,8.7429) (70,9.6279) (80,10.6725) (90,11.9195)
					};
					\addlegendentry{DP}
					\addplot[color=blue, mark=triangle*, mark size=2.5pt, line width=0.6pt]
					coordinates {
						(10,2.2617) (20,2.7978) (30,3.003) (40,3.2375) (50,3.5078)
						(60,3.8219) (70,4.1903) (80,4.6271) (90,5.1513)
					};
					\addlegendentry{CRM-Grid}
					\addplot[color=blue, mark=square*, mark size=1.5pt, line width=0.6pt]
					coordinates {
						(10,-0.7316) (20,-0.6804) (30,-0.6173) (40,-0.5392) (50,-0.4423)
						(60,-0.3212) (70,-0.1690) (80,0.0240) (90,0.2712)
					};
					\addlegendentry{CRM-IBR}
					\addplot[color=red, mark=x, mark size=2.5pt, line width=0.6pt]
					coordinates {
						(10,-0.7028) (20,-0.6466) (30,-0.5779) (40,-0.4936) (50,-0.3898)
						(60,-0.2609) (70,-0.0997) (80,0.1035) (90,0.3625)
					};
					\addlegendentry{ARM}
					\addplot[black, thin, dashed] coordinates {(0,0) (100,0)};
					\addplot[black, thin, dashed] coordinates {(80,-1000) (80,100)};
				\end{axis}
				\node[anchor=north, yshift=-29pt] at (current axis.south) {(c)};
			\end{tikzpicture}
			\captionsetup{labelformat=empty}
			\caption{}
			\label{fig:EG_3}
		\end{subfigure}
		\caption{Comparison of Distance Protection (DP), Classical Reactance Method (CRM), Advanced Reactance Method (ARM) during different grid topologies at bus B\textsubscript{1}: (a) only external grid, (b) external grid and GFL IBR, (c) only GFL IBR; $R_\mathrm{F}=\SI{2}{\Omega}$}
		\label{fig:Mit Netz}
	\end{figure}
	
	First, the external grid is connected to bus B\textsubscript{1}, while the GFL IBR is disconnected.
	Second, both, the external grid and the GFL IBR are connected to the bus.
	And third, the GFL IBR is connected to bus B\textsubscript{1}, while the external grid is disconnected.
	The fault resistance $R_\mathrm{F}$ is set to $\SI{2}{\Omega}$ while the fault location varies between 10\% and 90\%.
	Finally, the fault resistance is increased to $\SI{10}{\Omega}$ to investigate the robustness of all three protection algorithms against higher fault resistances.
	For the Classical Reactance Method two parameterizations are considered: On the one hand for the external grid (blue triangle in Fig.~\ref{fig:Mit Netz}), on the other hand for the GFL IBR (blue rectangle in Fig.~\ref{fig:Mit Netz}).
	This is intended to demonstrate what happens when the grid topology changes but the protection settings remain the same.
	
	For the first scenario, the external grid is connected to bus B\textsubscript{1}.
	As the short circuit power of both external grids is the same, the measurement deviations are expected to be very low.
	Fig.~\ref{fig:EG_1} shows the measurement deviation of all three protection algorithms.
	The Distance Protection, correctly parameterized Classical Reactance Method (CRM-Grid) and Advanced Reactance Method perform very well, with barely any deviation and nearly the same, while the wrongly parameterized Classical Reactance Method (CRM-IBR) shows a constant negative deviation.
	At the zone boundary of 80\% of the line length, Distance Protection shows a deviation of 0.20\%, while the correct Classical Reactance Method shows 0.02\% and the Advanced Reactance Method 0.04\%.
	The wrongly parameterized Classical Reactance Method shows a deviation of \text{-1.48\%}.
	
	For the second scenario, the GFL IBR is connected parallel to the external grid.
	Fig.~\ref{fig:EG_2} shows the measurement deviation of all three protection algorithms.
	The protection algorithms are affected by the FRT behavior of the GFL IBR.
	Still, Distance Protection, Classical Reactance Method with external grid parameters and Advanced Reactance Method show a low measurement deviation.
	At the zone boundary, Distance Protection shows a deviation of 0.43\%, while the Classical Reactance Method with external grid parameters shows 0.11\% and the Advanced Reactance Method 0.03\%.
	The Classical Reactance Method with the GFL IBR parameters shows a constant measurement deviation greater than -1\%.
	At the zone boundary, the deviation is \text{-1.30\%}.
	
	For the third scenario, the external grid is disconnected and only the GFL IBR is connected to bus B\textsubscript{1}.
	Fig.~\ref{fig:EG_3} shows the measurement deviation of all three protection algorithms.
	The Distance Protection shows great positive measurement deviations, while the correctly parameterized Classical Reactance Method (CRM-IBR) and the Advanced Reactance Method show good results.
	The wrong Classical Reactance Method (CRM-Grid) shows large positive measurement deviations.
	At the zone boundary, the Distance Protection shows a deviation of 10.7\%, the correctly parameterized Classical Reactance Method 0.02\%, the Advanced Reactance Method 0.10\% and the wrongly parameterized Classical Reactance Method 4.63\%.
	
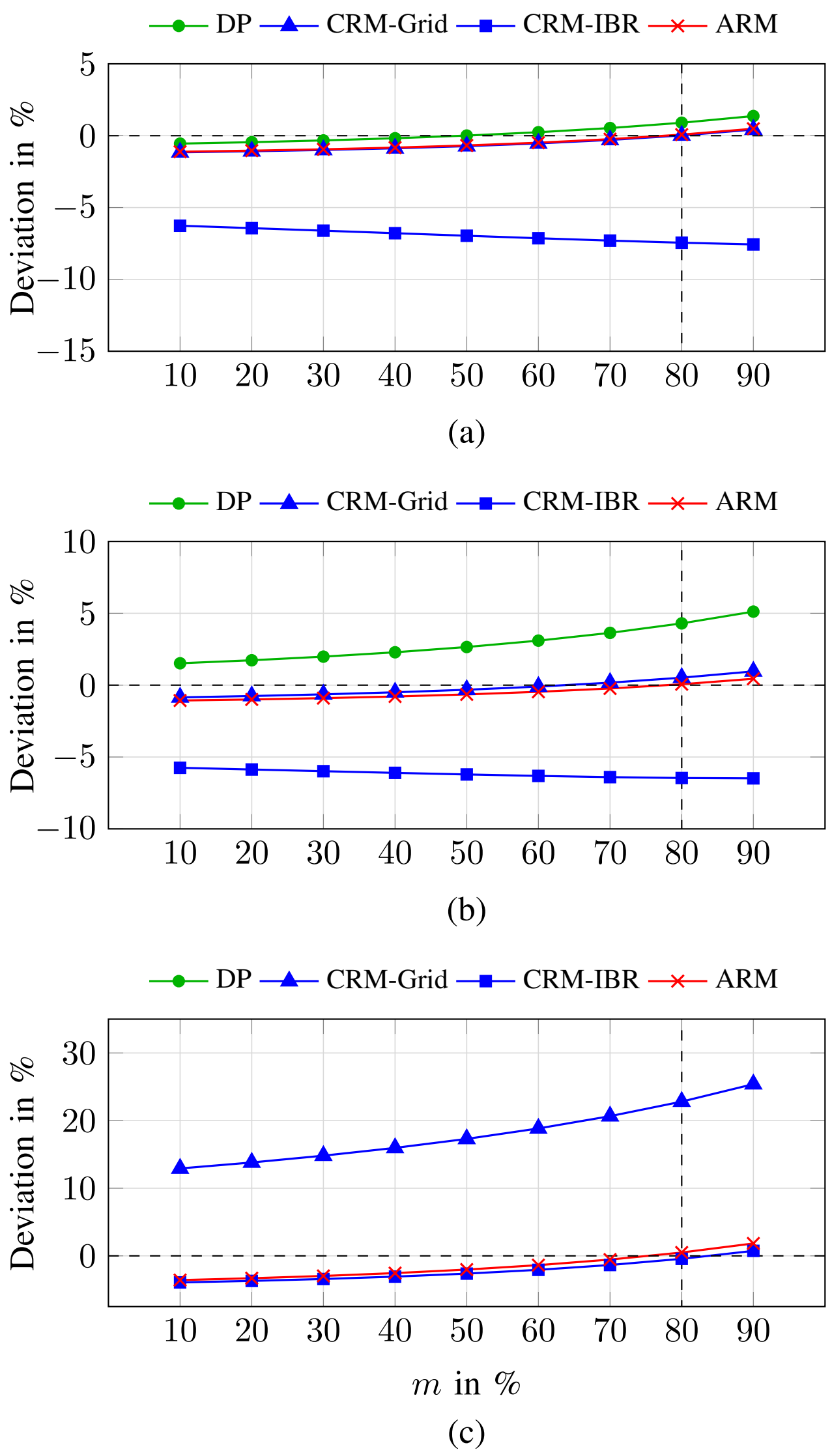
\begin{figure}[!t]
		\centering
		\captionsetup[subfigure]{skip=0pt}
		\setlength{\abovecaptionskip}{2pt}
		\setlength{\belowcaptionskip}{-5pt}
		
		\begin{subfigure}[t]{\columnwidth}
			\centering
			\begin{tikzpicture}
				\begin{axis}[
					width=\linewidth,
					height=4.5cm,
					xlabel={$m$ in \%},
					xlabel={},
					ylabel={Deviation in \%},
					ylabel style={yshift=-10pt},
					xtick={10,20,30,40,50,60,70,80,90},
					legend style={
						font=\footnotesize,
						at={(0.5,1.225)},
						anchor=north,
						legend columns=4,
						draw=none
					},
					grid=major,
					grid style={line width=0.3pt, draw=gray!30},
					xmin=0, xmax=100,
					ymin=-15, ymax=5,
					]
					\addplot[color=green!70!black, mark=*, mark size=1.5pt, line width=0.6pt]
					coordinates {
						(10,-0.5524) (20,-0.4525) (30,-0.3298) (40,-0.1788) (50,0.008)
						(60,0.2403) (70,0.5311) (80,0.8985) (90,1.3672)
					};
					\addlegendentry{DP}
					\addplot[color=blue, mark=triangle*, mark size=2.5pt, line width=0.6pt]
					coordinates {
						(10,-1.1563) (20,-1.0849) (30,-0.9935) (40,-0.877) (50,-0.7285)
						(60,-0.539) (70,-0.2961) (80,0.0171) (90,0.4239)
					};
					\addlegendentry{CRM-Grid}
					\addplot[color=blue, mark=square*, mark size=1.5pt, line width=0.6pt]
					coordinates {
						(10,-6.2644) (20,-6.435) (30,-6.6095) (40,-6.7863) (50,-6.9632)
						(60,-7.1365) (70,-7.3009) (80,-7.4482) (90,-7.566)
					};
					\addlegendentry{CRM-IBR}
					\addplot[color=red, mark=x, mark size=2.5pt, line width=0.6pt]
					coordinates {
						(10,-1.1156) (20,-1.0425) (30,-0.9493) (40,-0.8308) (50,-0.6801)
						(60,-0.4881) (70,-0.2424) (80,0.0738) (90,0.4841)
					};
					\addlegendentry{ARM}
					\addplot[black, thin, dashed] coordinates {(0,0) (100,0)};
					\addplot[black, thin, dashed] coordinates {(80,-1000) (80,100)};
					
				\end{axis}
				\node[anchor=north, yshift=-16pt] at (current axis.south) {(a)};
				
			\end{tikzpicture}
			\captionsetup{labelformat=empty}
			\caption{}
			\label{fig:EG_1_R2}
		\end{subfigure}
		\hfill
		\begin{subfigure}[t]{\columnwidth}
			\centering
			\begin{tikzpicture}
				\begin{axis}[
					width=\linewidth,
					height=4.5cm,
					xlabel={},
					ylabel={Deviation in \%},
					ylabel style={yshift=-10pt},
					xtick={10,20,30,40,50,60,70,80,90},
					legend style={
						font=\footnotesize,
						at={(0.5,1.225)},
						anchor=north,
						legend columns=4,
						draw=none
					},
					grid=major,
					grid style={line width=0.3pt, draw=gray!30},
					xmin=0, xmax=100,
					ymin=-10, ymax=10,
					]
					\addplot[color=green!70!black, mark=*, mark size=1.5pt, line width=0.6pt]
					coordinates {
						(10,1.5228) (20,1.7318) (30,1.9839) (40,2.2877) (50,2.6541)
						(60,3.0975) (70,3.6369) (80,4.298) (90,5.1164)
					};
					\addlegendentry{DP}
					\addplot[color=blue, mark=triangle*, mark size=2.5pt, line width=0.6pt]
					coordinates {
						(10,-0.8507) (20,-0.756) (30,-0.6396) (40,-0.4964) (50,-0.3195)
						(60,-0.1) (70,0.1742) (80,0.5196) (90,0.9591)
					};
					\addlegendentry{CRM-Grid}
					\addplot[color=blue, mark=square*, mark size=1.5pt, line width=0.6pt]
					coordinates {
						(10,-5.7503) (20,-5.869) (30,-5.9872) (40,-6.1031) (50,-6.2137)
						(60,-6.3149) (70,-6.4005) (80,-6.4616) (90,-6.4849)
					};
					\addlegendentry{CRM-IBR}
					\addplot[color=red, mark=x, mark size=2.5pt, line width=0.6pt]
					coordinates {
						(10,-1.0697) (20,-0.9962) (30,-0.9043) (40,-0.7893) (50,-0.6448)
						(60,-0.4626) (70,-0.2316) (80,0.0635) (90,0.4439)
					};
					\addlegendentry{ARM}
					\addplot[black, thin, dashed] coordinates {(0,0) (100,0)};
					\addplot[black, thin, dashed] coordinates {(80,-1000) (80,100)};
				\end{axis}
				\node[anchor=north, yshift=-16pt] at (current axis.south) {(b)};
			\end{tikzpicture}
			\captionsetup{labelformat=empty}
			\caption{}
			\label{fig:EG_2_R2}
		\end{subfigure}
		\hfill
		\begin{subfigure}[t]{\columnwidth}
			\centering
			\begin{tikzpicture}
				\begin{axis}[
					width=\linewidth,
					height=4.5cm,
					xlabel={$m$ in \%},
					ylabel={Deviation in \%},
					ylabel style={yshift=-10pt},
					xtick={10,20,30,40,50,60,70,80,90},
					legend style={
						font=\footnotesize,
						at={(0.5,1.225)},
						anchor=north,
						legend columns=4,
						draw=none
					},
					grid=major,
					grid style={line width=0.3pt, draw=gray!30},
					xmin=0, xmax=100,
					ymin=-7.5, ymax=35,
					]
					\addplot[color=green!70!black, mark=*, mark size=1.5pt, line width=0.6pt]
					coordinates {
						(10,72.769) (20,77.821) (30,83.469) (40,89.81) (50,96.961)
						(60,105.07) (70,114.3) (80,124.89) (90,137.08)
					};
					\addlegendentry{DP}
					\addplot[color=blue, mark=triangle*, mark size=2.5pt, line width=0.6pt]
					coordinates {
						(10,12.923) (20,13.807) (30,14.813) (40,15.965) (50,17.294)
						(60,18.84) (70,20.656) (80,22.809) (90,25.39)
					};
					\addlegendentry{CRM-Grid}
					\addplot[color=blue, mark=square*, mark size=1.5pt, line width=0.6pt]
					coordinates {
						(10,-3.9267) (20,-3.7028) (30,-3.4218) (40,-3.0691) (50,-2.6251)
						(60,-2.0641) (70,-1.3512) (80,-0.4389) (90,0.7384)
					};
					\addlegendentry{CRM-IBR}
					\addplot[color=red, mark=x, mark size=2.5pt, line width=0.6pt]
					coordinates {
						(10,-3.584) (20,-3.309) (30,-2.9696) (40,-2.5493) (50,-2.0266)
						(60,-1.3727) (70,-0.5494) (80,0.4958) (90,1.8354)
					};
					\addlegendentry{ARM}
					\addplot[black, thin, dashed] coordinates {(0,0) (100,0)};
					\addplot[black, thin, dashed] coordinates {(80,-1000) (80,100)};
				\end{axis}
				\node[anchor=north, yshift=-29pt] at (current axis.south) {(c)};
			\end{tikzpicture}
			\captionsetup{labelformat=empty}
			\caption{}
			\label{fig:EG_3_R2}
		\end{subfigure}
		\caption{Comparison of Distance Protection (DP) Classical Reactance Method, Advanced Reactance Method and  during different grid topologies at bus B\textsubscript{1}: (a) only external grid, (b) external grid and GFL IBR, (c) only GFL IBR; $R_\mathrm{F}=\SI{10}{\Omega}$}
		\label{fig:Mit Netz_R2}
	\end{figure}

	The fault resistance is now increased to $\SI{10}{\Omega}$.
	All the scenarios are simulated with the higher fault resistance.
	The results are shown in Fig.~\ref{fig:Mit Netz_R2}.
	In general, a higher fault resistance increases the deviation of the measured reactance in all scenarios.
	Still, it is evident that the Distance Protection and wrongly parameterized Classical Reactance Method are affected the most.
	The deviation of the Distance Protection increases but is still within an acceptable range for the first scenario, while the deviation increases for the second scenario, with the mixed infeed.
	For the third scenario, with the GFL IBR as only internal feeding source, the deviations reach very high values, ranging up to over 100\%.
	To present the other values clearly, the displayed range is not increased to these values.
	The deviation of the wrongly parameterized Classical Reactance Method increases dramatically due to the higher fault resistance.
	The correctly parameterized Classical Reactance Method and the Advanced Reactance Method both still show the best results for all scenarios with very limited deviations.
	
	\subsection{Evaluation of Deviation of External Grid Parameters}
	\label{C:Deviation of External grid}
	
	The robustness of the Advanced Reactance Method against changes in the external grid are evaluated.
	Switching actions or the enabling and disabling of generation units in the external grid may cause the short-circuit values to change.
	To simulate this, the impedance value of the external grid in the Advanced Reactance Method remain constant, while the actual values vary.
	The short-circuit power as well as the R/X ration are varied.
	The GFL IBR control consists of positive and negative sequence FRT.
	The fault resistance is $R_\mathrm{F}=\SI{2}{\Omega}$, while the fault location is varied between 10 and 90\% as before.
	
	Fig.~\ref{fig:Z_grid_variation} shows the results.
	The black values represent the reference, with the short circuit values of Table~\ref{tab:Evaluation}, which also represent the set values in the protection algorithm.
	The green values show changes in the short-circuit power (x = $\SI{3}{GVA}$, $\triangle$ = $\SI{20}{GVA}$).
	Overestimating the short-circuit power leads to an upwards shift and flattening of the reference line, while underestimation leads to the direct opposite behavior.
	The blue values show changes in the R/X ratio (x = $0.05$, $\triangle$ = $0.15$).
	An overestimation of the R/X-value leads to an downwards shift of the reference line.
	Moreover, the known behavior of the Advanced Reactance Method is a slightly negative deviation for faults lower the zone boundary and a positive deviation afterwards.
	This behavior changes by an overestimated R/X-value, as the deviation becomes more negative towards higher fault locations.
	Again, an underestimation of the R/X-value leads to the opposite behavior.
	
	Still, even at large deviation of the short-circuit power or R/X-ratio of the external grid, the deviations are small.
	At the zone boundary the deviations are 0.06\% for the reference, 0.43\% ($\SI{3}{GVA}$), -0.72\% ($\SI{40}{GVA}$), -1.18\% (0.05) and 1.28\% (0.15).
	
	\begin{figure}[!t]
		\begin{tikzpicture}
			\begin{axis}[
				width=\linewidth,
				height=4.5cm,
				xlabel={$m$ in \%},
				ylabel={Deviation in \%},
				ylabel style={yshift=-10pt},
				xtick={10,20,30,40,50,60,70,80,90},
				yticklabel style={},
				legend style={
					font=\footnotesize,
					at={(0.5,1.35)},
					anchor=north,
					legend columns=3,
					draw=none
				},
				grid=major,
				grid style={line width=0.3pt, draw=gray!30},
				xmin=0, xmax=100,
				ymin=-3.5, ymax=,
				]
				\addplot[color=black!90, mark=*, mark size=1.5pt, line width=0.6pt]
				coordinates {
					(10, -0.7028) (20, -0.6466) (30, -0.5779) (40, -0.4936) (50, -0.3898)
					(60, -0.2609) (70, -0.0997) (80, 0.1035) (90, 0.3625)
				};
				\addlegendentry{Reference}
				\addplot[color=green!70!black, mark=square*, mark size=1.5pt, line width=0.6pt]
				coordinates {
					(10, -0.2648) (20, -0.2457) (30, -0.2253) (40, -0.2034) (50, -0.1797)
					(60, -0.1543) (70, -0.1270) (80, -0.0976) (90, -0.0669)
				};
				\addlegendentry{$\SI{3}{GVA}$}
				\addplot[color=green!70!black, mark=triangle*, mark size=2.5pt, line width=0.6pt]
				coordinates {
					(10, -1.4138) (20, -1.3786) (30, -1.3224) (40, -1.2360) (50, -1.1052)
					(60, -0.9085) (70, -0.6106) (80, -0.1525) (90, 0.5701)
				};
				\addlegendentry{$\SI{20}{GVA}$}
				\addplot[color=blue, mark=x, mark size=2.5pt, line width=0.6pt]
				coordinates {
					(10, -0.9471) (20, -0.9732) (30, -1.0015) (40, -1.0320) (50, -1.0650)
					(60, -1.1008) (70, -1.1396) (80, -1.1816) (90, -1.2271)
				};
				\addlegendentry{$R/X=0.05$}
				\addplot[color=blue, mark=diamond*, mark size=2.5pt, line width=0.6pt]
				coordinates {
					(10, 0.4790) (20, 0.6436) (30, 0.8365) (40, 1.0644) (50, 1.3355)
					(60, 1.6606) (70, 2.0542) (80, 2.5354) (90, 3.1307)
				};
				\addlegendentry{$R/X=0.15$}
				\addplot[black, thin, dashed] coordinates {(0,0) (100,0)};
				\addplot[black, thin, dashed] coordinates {(80,-2) (80,2)};
				
			\end{axis}
		\end{tikzpicture}
		\caption{Classical Reactance Method for different values for the GFL IBR impedance with $R_\mathrm{F}=\SI{2}{\Omega}$}
		\label{fig:Z_grid_variation}
	\end{figure}
	
	\subsection{Evaluation of Parameterization of Reactance Method}
	\label{C:42_Parameterization of Classical Reactance Method}
	To evaluate the findings from \ref{C:31_Parameterization of the Classical Reactance Method in the Presence of GFL IBR}, four different parameterizations for the internal source impedance of the Classical Reactance Method are tested.
	First, to ten times the external grid impedance, as the short-circuit power of the external grid is ten times the rated power of the GFL IBR.
	Second, from the actual physical impedance of the GFL IBR, composed of the internal inverter and the transformer impedance.
	Third, to a pure reactance with the impedance based on the rated power of the GFL IBR.
	Fourth, to the impedance, described in \eqref{eq:3_GFL IBR NS Z}.
	Fig.~\ref{fig:x Grid_b} shows the results for the Classical Reactance Method for all four GFL IBR impedances for simulated fault with a fault resistance of $R_\mathrm{F}=\SI{2}{\Omega}$ at locations from 10\% to 90\% of the transmission line.
	The first two options show a large measurement deviation.
	This is due to the wrong impedance magnitude but moreover due to the wrong impedance angle.
	While the base reactance of the GFL IBR leads to good results, the proposed reactance $X_\mathrm{IBR}^{-*}$ leads to optimal results.
	For both values, the angle of the impedance is the same.
	The magnitude is better presented by the proposed variant.
	For faults at the zone boundary of $80\%$, the proposed variant leads to minimal deviation.
	The deviations for other fault locations are inherent in the Classical Reactance Method.
	
	In Table~\ref{tab:All}, all results at a fault location of 80\% are summarized.
	The results are discussed in the following chapter.
	
	\section{Discussion}
	\label{C:5_Conclusion}
	
	\begin{figure}[!t]
		\begin{tikzpicture}
			\begin{axis}[
				width=\linewidth,
				height=4.5cm,
				xlabel={$m$ in \%},
				ylabel={Deviation in \%},
				ylabel style={yshift=-10pt},
				xtick={10,20,30,40,50,60,70,80,90},
				yticklabel style={},
				legend style={
					font=\footnotesize,
					at={(0.5,1.225)},
					anchor=north,
					legend columns=4,
					draw=none
				},
				grid=major,
				grid style={line width=0.3pt, draw=gray!30},
				xmin=0, xmax=100,
				ymin=-1.5, ymax=6.5,
				]
				\addplot[color=green!70!black, mark=*, mark size=1.5pt, line width=0.6pt]
				coordinates {
					(10,3.1983) (20,3.4013) (30,3.631) (40,3.8925) (50,4.1926)
					(60,4.5398) (70,4.9453) (80,5.424) (90,5.9957)
				};
				\addlegendentry{$10\underline{Z}_\mathrm{grid}$}
				\addplot[color=black!90, mark=square*, mark size=1.5pt, line width=0.6pt]
				coordinates {
					(10,2.362) (20,2.5329) (30,2.7273) (40,2.9501) (50,3.2072)
					(60,3.5067) (70,3.8587) (80,4.2771) (90,4.7804)
				};
				\addlegendentry{$\underline{Z}_\mathrm{conv}$}
				\addplot[color=red, mark=triangle*, mark size=2.5pt, line width=0.6pt]
				coordinates {
					(10,-0.54) (20,-0.4814) (30,-0.4101) (40,-0.323) (50,-0.2161)
					(60,-0.0839) (70,0.0808) (80,0.2879) (90,0.5511)
				};
				\addlegendentry{${X}_\mathrm{IBR}$}
				\addplot[color=blue, mark=x, mark size=2.5pt, line width=0.6pt]
				coordinates {
					(10,-0.7316) (20,-0.6804) (30,-0.6173) (40,-0.5392) (50,-0.4423)
					(60,-0.3212) (70,-0.169) (80,0.024) (90,0.2712)
				};
				\addlegendentry{$X_\mathrm{IBR}^{-*}$}
				\addplot[black, thin, dashed] coordinates {(0,0) (100,0)};
				\addplot[black, thin, dashed] coordinates {(80,-1000) (80,100)};
			\end{axis}
		\end{tikzpicture}
		\caption{Classical Reactance Method for different values for the GFL IBR impedance with $R_\mathrm{F}=\SI{2}{\Omega}$}
		\label{fig:x Grid_b}
	\end{figure}
	
	Considering a protection relay on a bus, connected to a GFL IBR on the one hand and an external grid via a transmission line on the other hand, the Distance Protection does not work sufficiently.
	Table~\ref{tab:All} shows that the protection algorithm results greatly depend on the FRT behavior of the GFL IBR.
	Even when following typical grid code FRT requirements with the GFL IBR, the Distance Protection shows poor results because of the current source behavior and the low short-circuit power, compared to the external grid.
	Only in the presence of an equally strong external grid on the bus, the Distance Protection results are good.
	A protection algorithm must work properly, also if a connecting transmission line to an second external grid is disconnected.
	The findings lead to the conclusion that for such a grid topology, the basic Distance Protection algorithm, shown in \eqref{eq:2.1:LL Fault} violates the security, dependability and selectivity of the protection concept.
	A major advantage of the Distance Protection is the local measurement.
	This makes it attractive as backup protection for the differential protection, that needs communication.
	Failure of the differential protection can often be traced back to malfunction of the communication.
	The backup should therefore be without any communication.
	
	The Classical Reactance Method retains the advantage of the local measurement.
	Still, the algorithm needs more information about the grid, as the impedance of the local and external grid is necessary.
	Moreover, the algorithm is optimized for the zone boundary.
	This is the most critical part in Distance Protection.
	Still this assumption leads to an inherent measurement deviation for faults at other locations.
	For classical grids, with higher fault resistances, higher loadflow or different short-circuit powers of the infeeders, even with this disadvantage the Classical Reactance Method shows better results than the Distance Protection at all fault locations.
	Still, in the considered scenario the Classical Reactance Method shows ambiguous results.
	If the parameterization is correct and the GFL IBR follows negative sequence FRT, all results are good.
	When the grid topology changes, the protection algorithm must switch parameter groups to ensure good results.
	If this is not the case, it may lead to significant measurement deviations.
	As the algorithm does not work for GFL IBRs without negative sequence FRT, this must be known when designing the protection strategy.
	Without this negative sequence FRT, the Classical Reactance Method is not an option.
	
	The Advanced Reactance Method retains the advantage of local measurement.
	By incorporating the negative sequence voltage as a measurement variable, the method is independent on the local negative sequence impedance — a significant improvement over the Classical Reactance Method.
	However, it still requires the remote source impedance as a setting.
	Both methods are optimized for the zone boundary, meaning they deliver near-ideal accuracy when the fault location is correctly estimated.
	This results in minimal measurement deviation at the critical boundary point, which is the primary target.
	However, both methods exhibit an inherent deviation when the actual fault location differs from the assumed one, a known limitation.
	Crucially, the Advanced Reactance Method maintains its performance even when the GFL IBR suppresses negative sequence current during faults, a scenario in which the Classical Reactance Method is no longer applicable.
	This robustness makes the Advanced Reactance Method particularly suitable for modern grids with GFL IBRs, while preserving the same core advantages and limitations as the Classical Reactance Method.
	
	\begin{table*}[!t]
		\centering
		\caption{Results of the Distance Protection, Classical Reactance Method (correctly and wrongly parameterized) and Advanced Reactance Method for all considered scenarios with a fault location at the zone boundary at 80\% of the line. Results colored green if a deviation lower than 1\%, yellow between 1\% and 2.5\% and red higher than 2.5\%. NW = Not workable, due to the suppression of the negative sequence current.}
		\label{tab:All}
		\renewcommand{\arraystretch}{1.3}
		\begin{tabular}{l r r r r r r r r r r}
			\toprule
			\bfseries Protection Algorithm & \bfseries Fig.~\ref{fig:Eval_IBR_Q} & \bfseries Fig.~\ref{fig:Eval_IBR_P} & \bfseries Fig.~\ref{fig:Eval_IBR_Q_ONS} & \bfseries Fig.~\ref{fig:Eval_IBR_P_ONS} & \bfseries Fig.~\ref{fig:EG_1} & \bfseries Fig.~\ref{fig:EG_2} & \bfseries Fig.~\ref{fig:EG_3} & \bfseries Fig.~\ref{fig:EG_1_R2} & \bfseries Fig.~\ref{fig:EG_2_R2} & \bfseries Fig.~\ref{fig:EG_3_R2}\\
			\midrule
			Distance Protection & \cellcolor{red!20}10.67 & \cellcolor{red!20}-59.96 & \cellcolor{red!20}13.05 & \cellcolor{red!20}-64.71 & \cellcolor{green!20}0.20 & \cellcolor{green!20}0.43 & \cellcolor{red!20}10.67 & \cellcolor{green!20}0.38 & \cellcolor{yellow!35}1.09 & \cellcolor{red!20}30.33 \\
			
			Classical Reactance Method (correct) & \cellcolor{green!20}0.024 & \cellcolor{green!20}-0.29 & \cellcolor{red!20}NW & \cellcolor{red!20}NW & \cellcolor{green!20}0.02 & \cellcolor{green!20}0.11 & \cellcolor{green!20}0.02 & \cellcolor{green!20}0.02 & \cellcolor{green!20}0.19 & \cellcolor{green!20}-0.00 \\
			
			Classical Reactance Method (wrong) & - & - & - & - & \cellcolor{yellow!35}-1.48 & \cellcolor{yellow!35}-1.30 & \cellcolor{red!20}4.62 & \cellcolor{red!20}-2.97 & \cellcolor{red!20}-2.60 & \cellcolor{red!20}9.18 \\
			
			Advanced Reactance Method& \cellcolor{green!20}0.10 & \cellcolor{green!20}0.45 & \cellcolor{green!20}0.11 & \cellcolor{yellow!35}2.37 & \cellcolor{green!20}0.04 & \cellcolor{green!20}0.03 & \cellcolor{green!20}0.10 & \cellcolor{green!20}0.05 & \cellcolor{green!20}0.04 & \cellcolor{green!20}0.19 \\
			
			\bottomrule
		\end{tabular}
	\end{table*}
	
	Still, the results from Fig.~\ref{fig:Eval_IBR_P_ONS} show a vulnerability of the Advanced Reactance Method.
	The protection algorithm is designed for transmission grids with typical low $R/X$ ratios.
	For grid topologies with high $R/X$ ratios in certain current injection modes the denominator in \eqref{eq:3_X_meas_new} may approach zero and increase the sensitivity to measurement errors of the algorithm.
	For reasonable transmission grid values this does not happen.
	In Fig.~\ref{fig:Eval_IBR_P_ONS} a pure active power injection with suppression of the negative sequence current leads to a much higher sensitivity of the denominator in \eqref{eq:3_X_meas_new} than for all other cases.
	Thus, small measurement deviations or modeling approximations of the considered network, compared to reality lead to higher measurement deviations.
	However, this case represents the worst-case scenario with not following typical grid code requirements in both the positive and negative sequence and the deviation remains within acceptable limits.
	\\
	
	For the considered grid topology and a line protection relay on the GFL IBR bus, the authors recommend the following:
	\begin{itemize}		
		\item If the Advanced Reactance Method is implemented into the protection relay, it should be considered instead of or in parallel to the Distance Protection, either as backup for the Classical Reactance Method if there is no negative sequence current, or as full protection algorithm
		
		\item If the Advanced Reactance Method is not implemented into the protection relay, the GFL IBR FRT technique must be known. If it contains negative sequence current, the Classical Reactance Method should be considered, instead of or in parallel to the Distance Protection.
		
		\item The Classical Reactance Method must be parameterized correctly. Either as described in this paper, or—if there is more detailed information—according to the FRT of the GFL IBR
		
		\item Grid topology changes must be known and the Classical Reactance Method must switch parameter groups with this event
		
	\end{itemize}
	
	\section{Outlook}
	\label{C:6_Outlook}
	The results of the Advanced Reactance Methods show promising results.
	Still, there are tasks to be completed in the future.
	First, it must be meticulously analyzed under deviating grid parameters and line parameters.
	In \cite{polster2026faultlocation} this has already been taken into account for the Classical Reactance Method but not for the Advanced Reactance Method.
	Second, ground faults must be considered.
	Previous literature shows that the Classical Reactance Method works correctly for ground faults, if the zero sequence is taken into account.
	This sequence is not influenced by the GFL IBR, as the inverter is typically decoupled from the grid via a transformer.
	Therefore, this paper focuses on line-to-line faults.
	Nevertheless, a comprehensive protection algorithm should cover line-to-line faults and ground faults.
	Thus, the Advanced Reactance Method is tested for ground faults in future, as well with classical feeders, as with GFL IBRs.
	The authors see no disadvantage in the method, as the negative sequence can be taken into account, also for ground faults.
	The negative sequence data is often even more reliable than the zero sequence data.	
	
\section*{Acknowledgments}
\noindent The authors gratefully acknowledge funding by the German Federal Ministry of Research, Technology and Space (BMFTR) within the Kopernikus Project ENSURE ‘New ENergy grid StructURes for the German Energiewende’.
	
\bibliographystyle{unsrt}
\bibliography{2025_RM}

\begin{IEEEbiography}
	[{\includegraphics[width=1in,height=1.25in,clip,keepaspectratio]{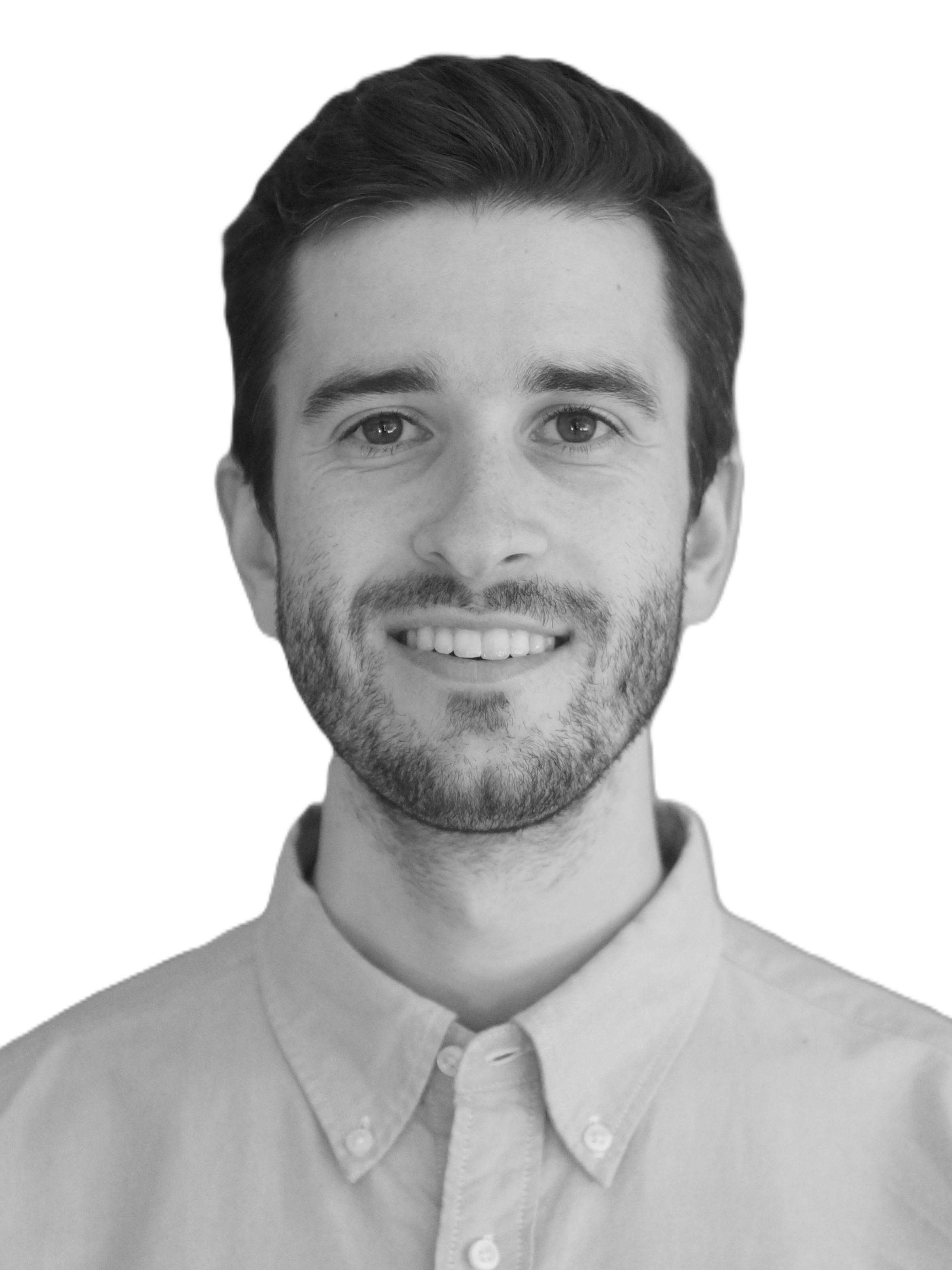}}]{Jonathan Loebel (Graduate Student Member IEEE)}
	Jonathan Loebel received the B.Sc. degree in Energy Engineering and the M.Sc. degree in Electrical Energy Engineering at the Friedrich-Alexander-Universität Erlangen-Nürnberg (FAU) in Erlangen, Germany in 2019 and 2022 respectively. Since 2022 he is a Research Associate with the Institute of Electrical Energy Systems at the FAU Erlangen. He works towards his PhD on grid protection under the influence of converter systems.
\end{IEEEbiography}

\begin{IEEEbiography}[{\includegraphics[width=1in,height=1.25in,clip,keepaspectratio]{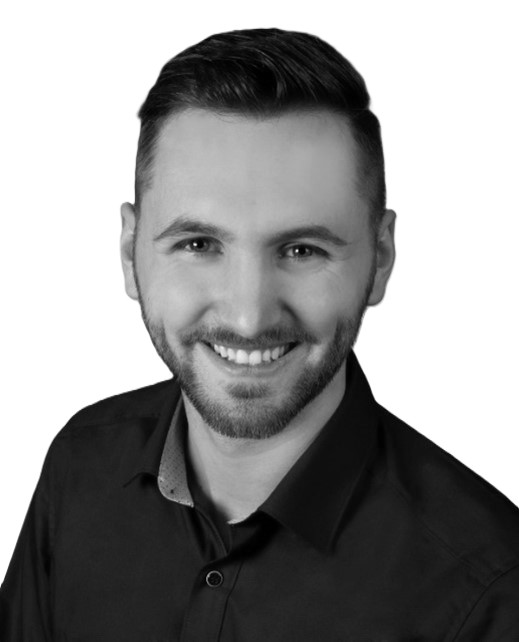}}]{Jannik Polster}
	Jannik Polster received the B.Sc. degree in Energy Systems and Energy Management from the University of Applied Sciences Ansbach in 2020 and the M.Sc. degree in Electrical Energy Engineering from Friedrich-Alexander-Universität Erlangen-Nürnberg (FAU) in 2025. Between 2020 and 2023, he worked as a technical project manager for PV systems and energy storage solutions. Since 2025, he is working with Siemens Energy as a System Design Engineer focusing on FACTS. 
\end{IEEEbiography}

\begin{IEEEbiography}[{\includegraphics[width=1in,height=1.25in,clip,keepaspectratio]{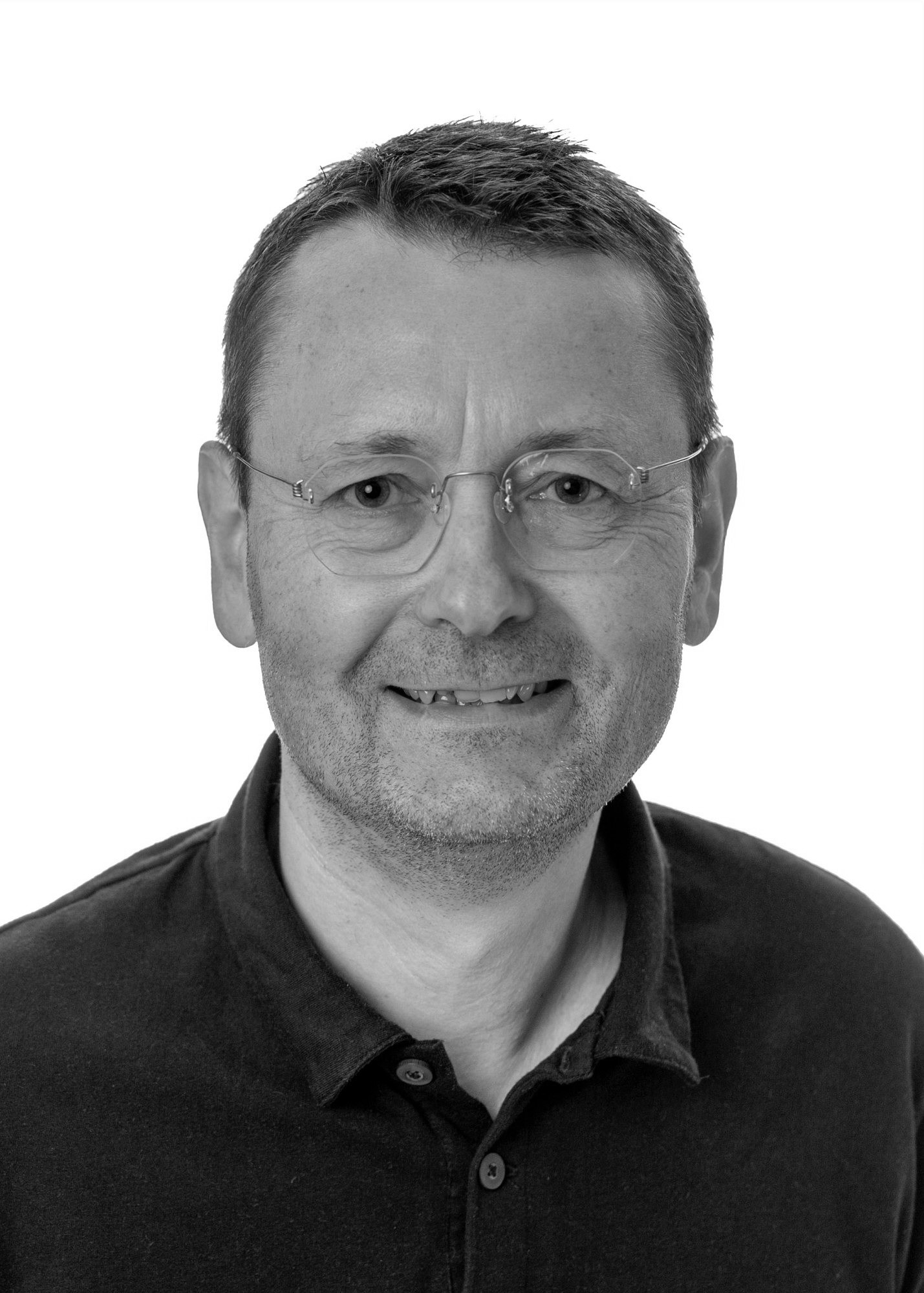}}]{Johann Jaeger (Member IEEE)}
	Johann Jaeger received the Dr.-Ing. degree 1996 in electrical engineering from the Friedrich-Alexander University Erlangen-Nürnberg (FAU). From 1996 he was with the Power System Planning Department at SIEMENS AG in Erlangen, Germany. He was working on different fields of FACTS devices, network planning and protections in worldwide projects. Since 2004 he is in charge of a full professorship for Electrical Power Systems at the FAU. He is member of VDE/ETG, IEEE and CIGRE as well as Convener or member of several national and international working groups and task forces. He is author or co-author of numerous papers, books and patents.
\end{IEEEbiography}

\end{document}